\documentstyle[12pt,aasms4]{article}
\begin{document}

\title {Limits on the Spatial Extent of AGN Measured 
with the Fine Guidance Sensors of the HST\footnote[1]{Based
on observations made with the NASA/ESA {\em Hubble Space Telescope},
obtained from the Space Telescope Science Institute, which is
operated by the Association of Universities for Research in Astronomy,
Inc., under NASA contract NAS 5-26555.}}

\author{Richard N. Hook}
\affil{Space Telescope European Coordinating Facility,
European Southern Observatory,
Karl-Schwarzschild-Str. 2, D-85748 Garching, Germany, rhook@eso.org}

\author{Ethan J. Schreier} 
\affil{Space Telescope Science Institute, 3700 San Martin Drive,\\
Baltimore, MD 21218, schreier@stsci.edu}

\author{George Miley}
\affil{Sterrewacht Leiden, Leiden University, P.O. Box 9513, 
NielsBohrweg 2, 2300 RA Leiden, The Netherlands, miley@strw.leidenuniv.nl}

\begin{abstract}
The optical structure of several AGN has been studied using the Fine
Guidance Sensors (FGS) on the Hubble Space Telescope (HST). The FGSs are
interferometric devices which can resolve structure on scales of 20 
milliarcsecs or less and hence have the potential to improve on the 
resolution attainable by HST's cameras. The FGSs produce interferometric 
fringes known as S-curves which are related to the intensity profile of 
the object on the sky.  These have been analyzed using a simple model for 
the  radial intensity distribution and strength of the underlying
background illumination of the observed objects.  Eight different observations 
of six different AGN have been analyzed.  No statistically significant 
differences from point sources are detected but significant upper 
limits of order 20 milliarcseconds are placed on any spatial 
extent.  Systematic effects limiting the resolution are discussed and 
some simple conclusions about the
physical size and luminosity densities of the emitting regions of the AGN
implied by the data are given.
\end{abstract}


\section{Introduction}

One of the most important properties of the Hubble Space Telescope (HST) is 
its ability to attain high spatial resolution limited by diffraction
rather than the properties of the atmosphere.  This resolution is
normally exploited via direct imaging by HST's cameras, potentially
enhanced by the application of restoration algorithms. The Fine Guidance 
Sensors (FGSs) of the HST offer an alternative method for studying
structure on the finest scales. The FGSs 
fold one half of the telescope's pupil onto the other and the resulting
interference fringes are recorded by photomultiplier tubes as the
5 arcsecond FGS field of view is scanned across a source. These fringes are
the basis of the HST's high precision tracking but can also be used
to study structure on scales down to 20 milliarcseconds or less. 
The FGSs are more sensitive than the cameras to the finest spatial
structure but produce data which are more difficult to interpret.

We have carried out a project to investigate the usefulness of the HST
FGSs for attaining ultra-high resolutions on extra-galactic objects, observing 
several bright AGN known to have radio structure on scales ranging from
several milliarcsecs to several tens of milliarcsecs. The goal was to search 
for and study similar scale structure in the optical. 

In this paper we present the results of these observations using the original
HST FGSs and their analyses using a simple model for detecting and 
quantifying extended structure. We apply statistical tests to decide whether 
the hypothesis that the object is a point source on an extended background 
is consistent with the data and if so, at what confidence level.  Simulations 
are used to derive limits on the detection thresholds for extensions to the 
central point-source. The systematic effects due to calibrations
incorrectly matched in both time and color are investigated as are
other systematic effects which limit the resolution we can attain.  
We find that all our datasets 
are consistent with point-like object intensity distributions.  The high 
spatial resolution of the FGS allows us to derive interesting limits
on the physical sizes of the optical emission regions, and on the scale and 
relative intensities of features which would have been seen. These results
delineate the region of phase-space where the FGSs are the best
observational tool for mapping at high resolution in the optical,
especially given the improved characteristics of the newly available
FGSs.

\section{Data from the FGS}
 
The HST uses three FGSs at the edge of the field of view of the telescope 
(see the FGS Instrument Handbook, Version 8, June 1999, STScI, for further 
discussion).  During normal use of the telescope, two of the FGSs are used 
to ``lock" onto guide stars to precisely point the telescope, while the
third is available to perform astrometry, independently scanning objects within
its field of view and measuring their positions.  Each FGS contains a pair 
of orthogonal interferometers which, when scanned across a point 
source, produce two S-shaped ``transfer functions."  The zero-point
crossings of the transfer functions determine the position of the 
source -- these nulls are used by the pointing control of the 
telescope -- while the morphology of the transfer functions are dependent on 
the brightness profile of the source.  Extended or multiple objects produce 
lower amplitude and more complicated transfer functions compared to those
from a point source. It may be shown that the
S-curve is related to the monochromatic intensity distribution of 
the object on the sky in the following way

\begin{equation}
S(\theta) = {{I_{A}-I_{B}} \over {I_{A}+I_{B}}} =
    \int_{-\Delta \phi/2}^{\Delta \phi/2} {F(\phi) \over F_{tot}}
    {{sin^2 ({D \pi (\theta - \phi) / \lambda})}
    \over {D \pi (\theta - \phi)/{ \lambda}}} d\phi
\end{equation}

where $S(\theta)$ is the value of the observed S-curve expressed as a
function of angle away from interferometric null,$I_{A}$ and $I_{B}$ are the
intensities measured by the photomultipliers, $D$ is the aperture of the HST,
$\lambda$ is the wavelength, $F(\phi)$ is the intensity distribution
of the object in the direction of the scan and $F_{tot}$ is the total
integrated flux. The observed S-curve can be regarded as the convolution 
of the normalized intensity profile of the object on the sky and the
S-curve resulting from a reference observation of a point-source on a 
negligible background. It follows from equation (1) that, in principle, the 
morphology of the source (in the direction of the scan) can be obtained 
by deconvolving the transfer function from the observed FGS interferometer
output.  A method for doing this has been described by Hershey (1992).  
Stronger backgrounds lead to lower amplitude S-curves and this effect
becomes important for faint sources such as those discussed in detail 
later in this paper.
To illustrate the general form of the FGS S-curves and how they change when
the object has small extent simulations were created and the results are
shown in Figure 1.

The primary use of the FGS in transfer function mode is for observations of 
binary stars (see, e.g. Franz et al. 1992), where separations, position 
angles, and relative intensities of close binaries are measured. 
Lattanzi et al. (1997) have also successfully used the FGS to measure the 
diameters of the disks of Mira variables. The observations discussed in 
this paper are the first to attempt to study extended structures in 
extragalactic sources with the FGS.

\section{The Observations and Data Reduction}

A list of datasets used in the analysis described below is given in 
Table 1.  All observations were made with FGS3 in the ``TRANS-MODE" (cf. FGS 
Instrument Handbook) to obtain transfer functions.  The initial
observations in this survey were carried out with three different filters: 
``PUPIL" (a filter-free stopped down aperture); ``CLEAR" (a broad-band 
filter with FWHM of 2340\AA\/ centered at 5830\AA); and ``YELLOW" (a
750\AA \/ FWHM filter centered at 5500\AA).  The original goal 
was to attempt to derive 
some color information from the filters, while the pupil stop, which apodizes
the outer $1/3$ of the telescope pupil, was used to restore fringe visibility
in the presence of spherical aberration.  Although in principle, the 
interferometers should not be affected by the symmetric aberration in the 
primary mirror, small internal FGS misalignments interact with the spherical 
aberration to impair FGS performance. All later observations, forming the
majority of the data presented here, used the pupil stop only.

Each observation consisted of multiple scans across the object.
The two orthogonal interferometers of the FGS were oriented at 45 degrees 
to the the scan direction.  Each scan was typically 2.4" long, centered 
on the object, with a nominal 0.3 mas step size. 

Routine data reduction software developed by the FGS group at STScI,
working with the HST Astrometry team, was used to unpack, inspect, smooth, 
and merge the data sets (see, e.g., HST Data Handbook, Part X, February, 
1994, STScI).  Individual scans which on visual inspection showed no 
obvious transfer function or displayed 
pathological (non-physical) characteristics due to spacecraft jitter, 
were not considered further.  On the average, 75\% of the scans were 
considered ``good."  The data from these scans were then co-added,
separately for each filter and for each coordinate axis.  The standard merging 
software was used, which first determines a ``zero phase" for each scan 
by measuring the zero crossing of the transfer function (observed in the 
smoothed data), then calculates the relative offsets of the observed
transfer functions from that of a reference scan (typically the first 
scan of the set), and then co-adds the raw (unsmoothed) data from the 
individual scans, using the calculated offsets. For the data from the faint 
objects discussed here each individual
scan has significant photon noise. This limits the accuracy of the offset
determination and degrades the resolution which can finally be attained. This
source of systematic error is discussed below in section 5. 

It was apparent that the pupil data were the ``cleanest" in the sense that 
the amplitudes of the transfer functions were greatest and the shapes were
smoothest and most closely resembled expected transfer function shapes.  
This is expected because the full aperture, which accepts a larger portion
of the spherically aberrated optical beam, is more susceptible to internal
FGS misalignments. In order to have a uniform set of datasets for many objects,
which could be adequately calibrated with the limited set of
calibration curves at our disposal, we decided to restrict the analyses
presented here to those datasets obtained using the pupil stop.
Unfortunately, use of the pupil did reduce the photon flux, and thus the 
number of sources we could study with this technique, as well as
the angular resolution which could be achieved.

In order to analyze these curves, we require for reference the response of
the system to a known point-source on a negligible background, effectively
the S-curve ``point spread function" (PSF).  The FGS is routinely
scheduled to observe standard stars of different colors, through
different filters, to provide such references.  We selected as reference
S-curves for our analyses observations of stars observed through the pupil
aperture as close as possible in time to our observations, as the
FGS3 S-curves are known to change over the course of time.  Ideally a 
reference star with color comparable to that of the object should be 
selected, but this can only be approximated for AGN. It is also not always 
possible to find reference curves taken less than about 200 days from a 
given observation. The systematic effects due to the inadequacies of the 
reference curves in color and time are discussed below in section 5. 
A list of all calibration curves used in this paper is given as Table 2.

\section{Model-fitting the Observed S-curves} 

The merged transfer functions for several objects observed in 
our sample appeared clearly inconsistent with what would be 
expected from an ideal point source.  In particular, the 
amplitudes of the S-curves were less (suggesting that the 
background was a significant fraction of the object intensity) 
and the widths appeared greater (suggesting extent) than expected for 
point sources. To understand the significance of these 
differences, we created model object intensity profiles, generated 
sets of simulated S-curves from them, and compared these to the 
observations. In this section we use the 3C279 dataset 
f0wj0602m to illustrate the methods used to assess the 
statistical significance of the structures seen in the S-curves. 
The other data sets were analyzed in the same way and the results
are presented below.

The objects are modeled using normalized Gaussian intrinsic 
intensity profiles, superimposed on flat backgrounds.  The 
two-parameter set of functions we used for the intensity 
distribution in each of ``x" and ``y" is thus of the form:

\begin{equation}
F = B + H e^{ - \theta^2 /{2\sigma^2}}
\end{equation}

where B and H are constrained so that the sum over all values of F is 1.0, 
$\theta$ is the angle with respect to the FGS interferometry null 
(normally expressed in milliarcseconds) and $\sigma$ is expressed in the 
same units.  The normalization of this function is important and may be 
derived easily from the theory of the FGSs (see eq. [1]).  This function 
$F$ is a model for the intensity profile of the object in one direction, 
summed over the 5 arcsec width of the FGS instantaneous field of view in 
the perpendicular direction.  The model S-curve itself is then created by 
convolving the model source profile F with the appropriate calibration 
S-curve - the ``PSF" - as listed in Table 1. 

The observation data sets were reduced using the standard FGS software 
referenced earlier. The noisy, un-smoothed data products rather than the 
smoothed versions were used (both are created by the standard FGS
software), 
as we wished to retain the noise properties of the data.  For the
calibrator S-curves, the smoothed version was used, re-sampled to match 
the scale and range of the observed data.  In both cases a 512 sample
subset of the data around the central S-curve feature was used to make the
convolutions more efficient. This reduced range does not affect the
results, 
as there are no significant features further out in these data.

The observed data sets were compared with the models (convolutions of the 
intensity profiles with calibration S-curves) for a range of values of 
$\sigma$ and $B$, and the sum of squared residuals was computed.  To 
estimate the statistical significance of the results, we measure the 
noise in the outer parts of the S-curve, where the mean is zero, and 
assume it to be approximately constant. The reduced $\chi^2$ for the 
fits is then calculated in the normal way
\begin{equation}
\chi^2_{red} = {{\sum { (v_{dat}-v_{model})^2 / var}} \over
n_{free}}
\end{equation}
where $v_{dat}$ and $v_{model}$ are the values of the data and model at a
given
position, $n_{free}$ is the number of degrees of freedom, in this
case this is the number of data points (512) minus the two free fit
parameters (the width of the Gaussian and the fraction of energy in
the background) and $var$ is the measured variance of the noise, assumed
to be constant throughout the S-curve. It is assumed that there is no
correlation between the data points; this assumption will be discussed 
further below.

Confidence levels are plotted in Figure 2 for the two-parameter fits 
to the 3C279 data sets (f0wj0602m).  For a $\chi^2$ distribution with 510
degrees of freedom a reduced $\chi^2$ of 1.072 corresponds to the 25\%
confidence level, 1.123 to 5\% and 1.161 to 1\%. 
The reduced $\chi^2$ values of the good fits are slightly less than $1.0$,
implying that the noise as estimated from the outer parts of the S-curve
is slightly too large, in agreement with visual inspection; the data are not
``over-fitted,'' as the curve is smooth on large scales and cannot fit any
but the largest features in the S-curve. Estimating the variance 
from the outer parts of the
S-curve gives a spread around the truth with the effect of shifting the
confidence contours in or out.  We use the $F$ test (below), based on
ratios of $\chi^2$ values, as it is not affected by small variations of
the variance measure. 

For statistical tests to be valid, we need to know the magnitude of the
noise, its distribution, and whether there are correlations present which
would reduce the effective number of degrees of freedom. To assess these
properties, we analyzed a ``noise dataset'' produced from the difference
between the fitted and observed S-curves for 3C279 (the ``y'' data were
used but ``x'' would be equally good). Again, because the fitted curve is
smooth on all but the largest scales, this will be a good guide to the
fine structure of the noise.  The histogram of this noise dataset appeared
approximately Gaussian with zero mean, a width (i.e. the $\sigma$ of the
noise) of 0.044, and no obvious asymmetry.  Figure 3 shows the power
spectrum of the noise which appears to be approximately ``white''.
It is plotted with the zero-frequency bin at the center and is hence symmetrical
about this point.
Finally, the correlations within the noise were assessed by block-averaging the
data in 2, 4 and 8 sample bins and computing the standard deviation in each. If
there are no correlations on these scales, the values should drop by
$\sqrt{2}$ each time. The actual ratios found were 1.50,1.47 and 1.36,
confirming that the noise may be regarded as Gaussian and independent,
with 512 degrees of freedom.

As seen from Figure 2, there is a large region of acceptable parameter
space. The best fit requires a small extension (FWHM about 15mas) with a
background of about 24\%. However, a point source with a background about
28\% of the total intensity, is also fully acceptable; the required
background is consistent with the level expected from on object of this
magnitude. The hypothesis that there is no background, and that the
reduction in S-curve amplitude is due to a broad nucleus, is strongly
rejected at a significance level of 99.9\%. 

We adopt two approaches to further assess the significance of the small
extension indicated by the minimum $\chi^2$ fit.  First, we ask whether
the improvement in the fit due to the addition of an extra parameter to the
model (the extent of the object) is justified by the data.  We compare the
$\chi^2$ of the null hypothesis ``the object is a point'' with that
obtained by including a non-zero width as well. The ratio of these two
$\chi^2$ values should follow the $F$ distribution, allowing us to put a
confidence limit on the extension. The use of the $F$ test also removes
the uncertainty in the measurement of the variance, as we are comparing a
ratio of reduced $\chi^2$ values.

From the 3C279 S-curves we have
\begin{equation}
F = {{\chi^2_{extended}} \over { \chi^2_{point}}} = 0.9807/0.9682 =
1.013
\end{equation}
for x and
\begin{equation}
F = {{\chi^2_{extended}} \over { \chi^2_{point}}} = 0.9892/0.9573 =
1.033
\end{equation}
for y, where there are 511 degrees of freedom for the point-source fit
and 510 for the two parameter fit. For an $F$ distribution with 510 and
511 degrees of freedom, a 10\% confidence level is reached at $F_{510,511,0.9}
= 1.12$. In other words, such a high value is to be expected by chance in
10\% of cases if the null hypothesis were true.  Although this application
of the $F$ statistic is not rigorously correct, as the fitted function is
not linear in the width of the Gaussian, we think it is clear that we
cannot reject the null hypothesis and hence cannot put any trust in an 
extension, when the derived $F$ values are so close to $1.0$. 

In a second approach, we generate Monte Carlo simulations using the same
two parameter model, selecting the width and fractional background
strength at random, convolving the models with the same calibrator 
S-curve, and adding Gaussian noise at levels comparable to that 
present in two data sets of interest.  The results were analyzed 
in exactly the same way as above to measure the width and 
background. Figure 4 compares the measured widths with those
input to the simulations for two noise levels. For the noisier
simulations on the left, which have noise comparable to that 
seen in the 3C279
S-curves, there is reasonable agreement for extents
($\sigma$) greater than about $10$ mas.  In a higher signal
to noise case on the right, where the noise is comparable to that
seen in the NGC4151 data set, extents down to about $5$ mas can
be reliably detected. These plots provide a guide to
whether or not an extension can be detected at a specified noise 
level when other sources of systematic error are negligible. We
quantify the statistical significance of the extent 
determination by again looking at the $F$ statistic for each of 
these S-curves, as plotted in Figure 5.  The pluses indicate fits which are
non-point-like at the 99\% confidence level, the stars 
fits which have $F$ values insufficiently different from $1.0$ to 
allow rejection of the null hypothesis (i.e. point-source) at the 
25\% level. The triangles are intermediate cases. 

It is clear from the left-hand plots that $\sigma$ values less 
than about 12mas (FWHM=28mas) will not be detectable at 
statistically significant levels in S-curves having noise levels 
comparable to the 3C279 data. Hence the extension of $\sigma=6$mas
(FWMH=14mas) we obtained for the 3C279 ``x" S-curve is not 
significant. Similarly the smaller extension measured for the
``y'' S-curve of the NGC4151 data described later in the paper
is also not statistically significant. In this case, that of the
best signal to noise of our sample, the smallest
statistically significant extent ($\sigma$) detectable is of
order $5$mas. For data of lower 
signal to noise ratio, including most of the other data sets 
studied here, the minimum extension which can be detected will 
clearly be larger than that for 3C279. Figures 4 and 5 also show
that the minimum detectable extension falls as the fraction of
energy in the background increases. This is to be expected as
the amplitude of the S-curve, compared to the constant noise, is
also falling in this case.

These simulations allow us to place limits on the spatial extents of
the objects in our study which would be detected at different noise levels
under the assumption that other systematic effects are negligible. 
Unfortunately the combined effects of the spherical aberration of the
HST, small misalignments within FGS3 itself, variations in the
fringe shapes with time and the impossibility of accurately coadding shifted
fringes with very low signal restrict the resolution which can be obtained.
These limits are discussed below.

\section{Systematic Effects}

The above analysis has assumed the calibration reference S-curves are an
accurate representation of what would be seen if the observed object were
a point on a negligible background. It is also assumed that the observed
S-curves can be regarded as the convolution of the normalized object intensity
distribution with the calibration S-curve. Unfortunately neither of these
assumptions can be regarded as completely valid for S-curves of faint objects
observed with the FGS3 interferometer. 
In practice, S-curves depend on the color of the object and are also known to 
vary in time.  In addition, the accuracy of aligning and coadding the multiple
scans across a faint object is limited by the photon noise of each scan. 
In this section we discuss how these systematic effects limit our results.

The effect of color on the S-curve follows simply from the definition of
the transfer function: equation (1) contains the dependence of the S-curve
on color.  For a monochromatic S-curve, changing the wavelength has the
effect of scaling the S-curve with respect to the $\theta$ axis. This is a
simple effect which can be modeled for an object of known spectral-energy
distribution by also convolving over wavelength.  We generated two simulated
S-curves based on two extremes -- a very blue spectrum and a very
red spectrum. We then used the red S-curve as a calibrator to
analyze the blue using the procedures discussed above. The color
difference, far larger than any realistic case, is found to introduce 
a ripple in the residuals of the best fit but no measurable extent. 

We also took the spectrum of NGC4151 and multiplied by a simulated FGS spectral
response and used this to generate a simulated S-curve corresponding to
a point source having the appropriate SED. We then made another simulated
S-curve with spectral weighting appropriate for the calibration star Upgren 69.
To test the color effect on our results we analyzed this simulated
NGC4151 observation using the simulated Upgren 69 S-curve as calibrator. The
resultant fit showed low amplitude residuals but no detectable bias in the
measurement of extent or background, both of which were zero at the resolution
of the analysis.

The time variation of the S-curves is much less well understood than
the color dependence. It presumably depends on time-dependent changes in 
the geometry of the fine guidance sensors. Experience with FGS3 data on 
double stars and the disks of Mira stars has shown that the variations, 
which are particularly marked for scans in ``x'', limit resolution
to approximately 20mas. Unfortunately, there are at present insufficient data 
to allow predictions of this effect. This, and other sources of systematic 
effects, are discussed in the FGS Instrument Handbook. 

To empirically estimate the effects of time-variant systematics on the 
analyses of our data sets as described above, we re-processed a high-signal 
to noise case (f2m40301, 3C 273) using a set of different reference 
calibrations S-curves listed in Table 2 which were taken after the first 
HST servicing mission. These cover a period from 280 days before the 
observation to 740 days afterwards. The last calibration set listed, below 
the line, is a bluer star.  The resultant fits are given in Table 3.

There is significant variation of quality of fit when different calibration
curves are used. As expected, the reference curve closest in time to the
observation (f2vc0201m) does seem to give a better fit than those obtained
a long period either before or after, but the differences are small.  All
fits give very similar numerical values for both the fitted parameters, on
both axes, within the formal uncertainties as calculated before. In this
case we find consistently that both the background level and the width 
are negligible.

A final important source of systematic error is introduced by the reduction
of the multiple FGS scans. Each of these has very low signal-to-noise for
the faint objects considered here and each has a random, unpredictable shift
relative to the others. To coadd these scans, as described in section 3 above,
it is necessary to measure their relative shifts. This process cannot be done
accurately when noise is very significant and this process inevitably
degrades the resolution attainable.

The systematic effects described above combine to limit the
resolution, and hence the minimal detectable object extent, to approximately
20mas, and are greater than the more fundamental limits due to the pure photon
noise in the observed S-curves. This limit applies to the FGS3 interferometer
when used in pupil mode. The newer FGS1R interferometer, installed during the
second HST servicing mission, has been shown to have much more stable 
S-curves and also doesn't require the use of the pupil stop. Hence it should 
have much improved performance for this kind of work.

\section{Results for Individual Objects}

All the objects listed in Table 1 were fitted to the two-parameter family
of models described above and the minimum $\chi^2$ fit results are
shown in Table 4. The data are shown in Figures 6 to 13. In all
figures, the best fit is shown along with the data in the upper plot 
and the residuals are plotted below, on the same scale. 
Only the 512 points used in the analysis are plotted. In each subsection
below, we summarize the results for each object.

\subsection{3C279 (Figures 6 \& 7)}
3C279 is an extended double radio source which includes a compact core
and a jet which extends about 5 arcsec (de Pater \& Perley 1983)
as well as structure observed with VLBI on scales down to 0.1 mas
(B\aa \aa th et al. 1992). The jet of 3C279 was one of the first
known examples of superluminal motion.  3C279 is
associated with a QSO at redshift 0.538
(Sandage \& Wyndham 1965).  It is a luminous X-ray and gamma-ray source,
and is highly variable at all wavelengths.  In the optical it is highly
polarized, with V magnitude ranging from 15th to 17th magnitude during
recent years.

There are two observations of this object and the earlier ones were
used in the above examples of modeling methods. The data are of reasonable 
quality and the fits are good. The data are consistent with
a point-source on a moderate background, consistent with the object's
brightness.

\subsection{NGC1275 (Figure 8)}
These data are of very low signal to noise and the amplitude is very
low because of the relatively high background level from this
bright galaxy. The fits are reasonable and consistent with a point-source.

\subsection{3C345 (Figures 9 \& 10)}
Both datasets for this object have low signal to noise. The fits are
good and are consistent with point-sources.

\subsection{NGC4151 (Figure 11)}
These S-curves have excellent signal to noise, the best of all those
studied here. Unfortunately the fits are not very good as is shown in
the $\chi^2$ values. It seems most likely that this is due to an
inadequately matched reference S-curve due to variations with time of
the S-curve form. This dataset is considered further in the Section 6 above.

\subsection{3C273 (Figure 12)}
These curves also have excellent signal to noise and the fits are
reasonable although, like NGC4151, it appears that better calibrations
would allow more information to be extracted. In terms of amplitude and
width these curves are very similar to the calibration indicating that the
relative background is low and that the object is, to first approximation,
a point. The X S-curve , which is less stable than in Y, shows some
systematic differences. Possible explanations for such differences
were discussed in section 5 above.
The Y S-curves fit is closer but there is also a systematic difference.

\subsection{M87 (Figure 13)}
These S-curves have very small amplitudes because of the strong background
from the underlying elliptical galaxy. The fits are consistent with a
point source but the signal to noise ratio is very low.

\section{Discussion}

All the bright AGN which have been studied using the HST FGS in TRANS mode
can be modeled well by assuming that the true intensity
distribution of the object is a point source on a background. 
Statistical tests show that the very small measured widths
are not significant.  The levels of background
measured are consistent with what is expected: bright point-like objects
(eg, 3C273) have negligible background and objects which are the cores of
bright, large galaxies (eg, M87 and NGC4151) have strong backgrounds. For
faint objects the instrumental background is detected and appears at the
expected strength (eg, 3C279).

An assessment of the systematic effects limiting the resolution of the FGS3
interferometer imply that we would not expect to detect extents (FWHM) less
than 20mas ($\sigma=9$ mas). These effects are dominated by the instability of
the FGS3 interferometer in combination with the spherically
aberrated telescope optics and the difficulty of accurately coadding individual
scans of faint objects.  

We may use our results to estimate the physical extents of the 
emitting regions and the luminosity densities which these imply. 
Table 5 tabulates these quantities for those objects with statistically
significant upper limits. We assume a spherical, uniform, 
optically-thin emitting region, $H_0 = 65km s^{-1} Mpc^{-1}$,
and that the upper limit to the radius of the sphere is the 
measured $\sigma$. The luminosities are calculated from the 
$V$ magnitude of the objects and the sun with no consideration of color 
effects. These objects are variable and hence approximate mean values are 
used for the $V$ magnitude. For the two bright objects we use the limit of
$\sigma=9$ mas as deduced from a study of the systematics. For the fainter
3C279 the noise of the data itself becomes more important and we use the
limit $\sigma=12$ mas. For the brightest objects in out sample the FGS3 
observations are limited by the systematic effects described in detail 
in section 5 above.

Because it is both the closest object in our study and has the S-curves
with the highest signal to noise ratio, the NGC4151 data yield the smallest
physical limit on the size of the nuclear emission region. This object
is the closest and most studied Seyfert I galaxy and has been modeled in
detail in the framework of the unified view of AGN by Casidy and Raine 
(1997). The broad line region (BLR) of this object varies dramatically on
time-scales of weeks and must have an angular extent below 1mas. The
narrow line region (NLR) further out extends over hundreds of parsecs and
has been studied in detail from the ground and by imaging with HST. Our
FGS observations are not sensitive to these extended features on arcsecond
scales; they are included in the measured background. We expect any structure
we see on scales of 10mas to come from material around the BLR which is
scattering the intense nuclear radiation. If such scattering was bright
enough to effectively broaden the nuclear intensity profile in the optical
we would detect such extent. Because of our observations are consistent
with a point source it appears likely that such scattering is overwhelmed
by the direct view of the BLR or that such a broadening occurs only
closer to the BLR.

\section{Conclusions}
The original HST FGS3 interferometer has been shown to be capable of 
measuring extents down to about $\sigma=9$ mas -- on the order of a parsec 
in nearer AGNs. This surpasses the capabilities of 
HST PC camera by close to an order of magnitude. The upper limit is due to a
combination of photon statistics and systematic effects not predictable in the
pre-refurbishment FGS. 
If these objects were re-observed using the post-refurbishment FGS1R these
systematics would be dramatically reduced and the attainable performance would
become limited by the photon statistics of the data. This would lower the
upper limits which could be placed on the extent of the AGN emitting regions to
the values given in Table 6.  The refurbished FGS1R, with
S-curves which are more stable and closer to the theoretically perfect
form, and which do not require the pupil stop which degrades both throughput and
resolution, will allow detection of structure on scales of 5mas or smaller
on the brighter AGNs.

\acknowledgements
We are grateful to many people who contributed to this project
over many years. Doris Daou and Nicola Caon gave invaluable help
with the data reduction. Mario Lattanzi, Pierre Bely, and Sherie Holfeltz
provided much useful advice and encouragement in many discussions.
We especially thank Ed Nelan for his simulations, his advice about the FGS
systematics, and his careful reading of the manuscript.

We would like to acknowledge support from STScI grants GO-02443.01-87A 
and GO-06578.01-95A.

\clearpage

\begin{figure}
\plotone{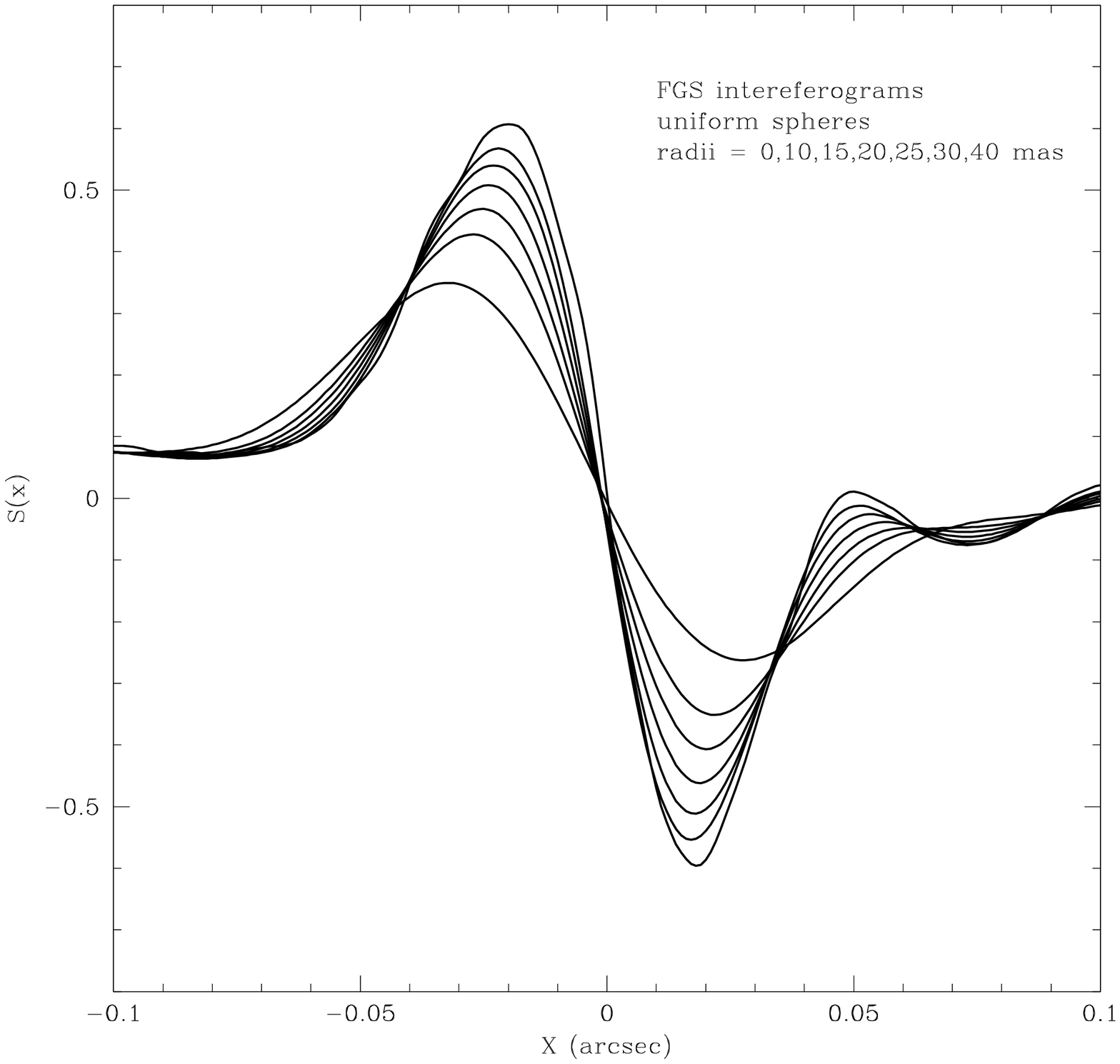}
\figcaption{Predicted HST FGS S-curves for uniform, optically thin spherical
sources of the radii shown.}
\label{Figure 1}
\end{figure}

\begin{figure}
\plottwo{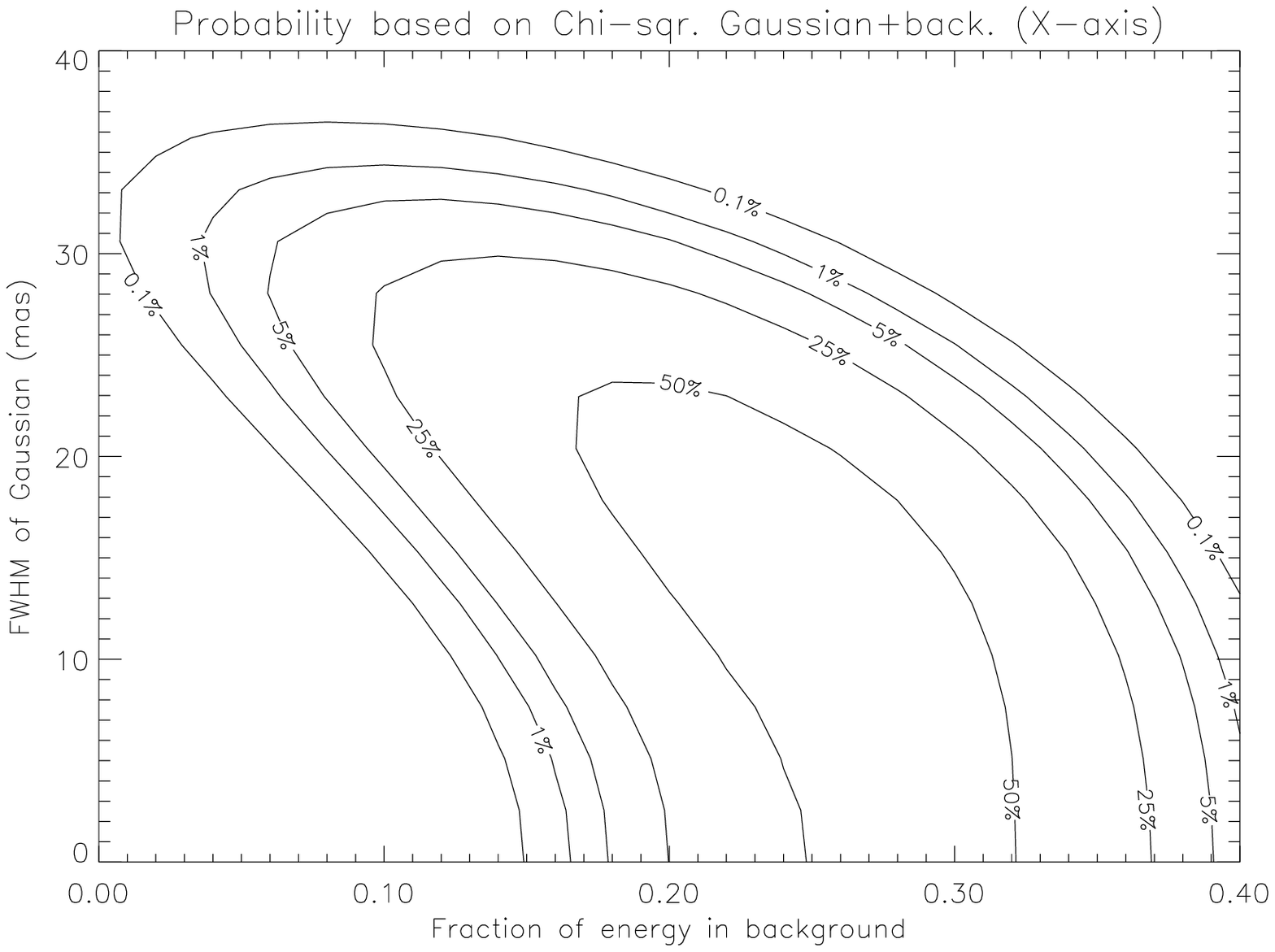}{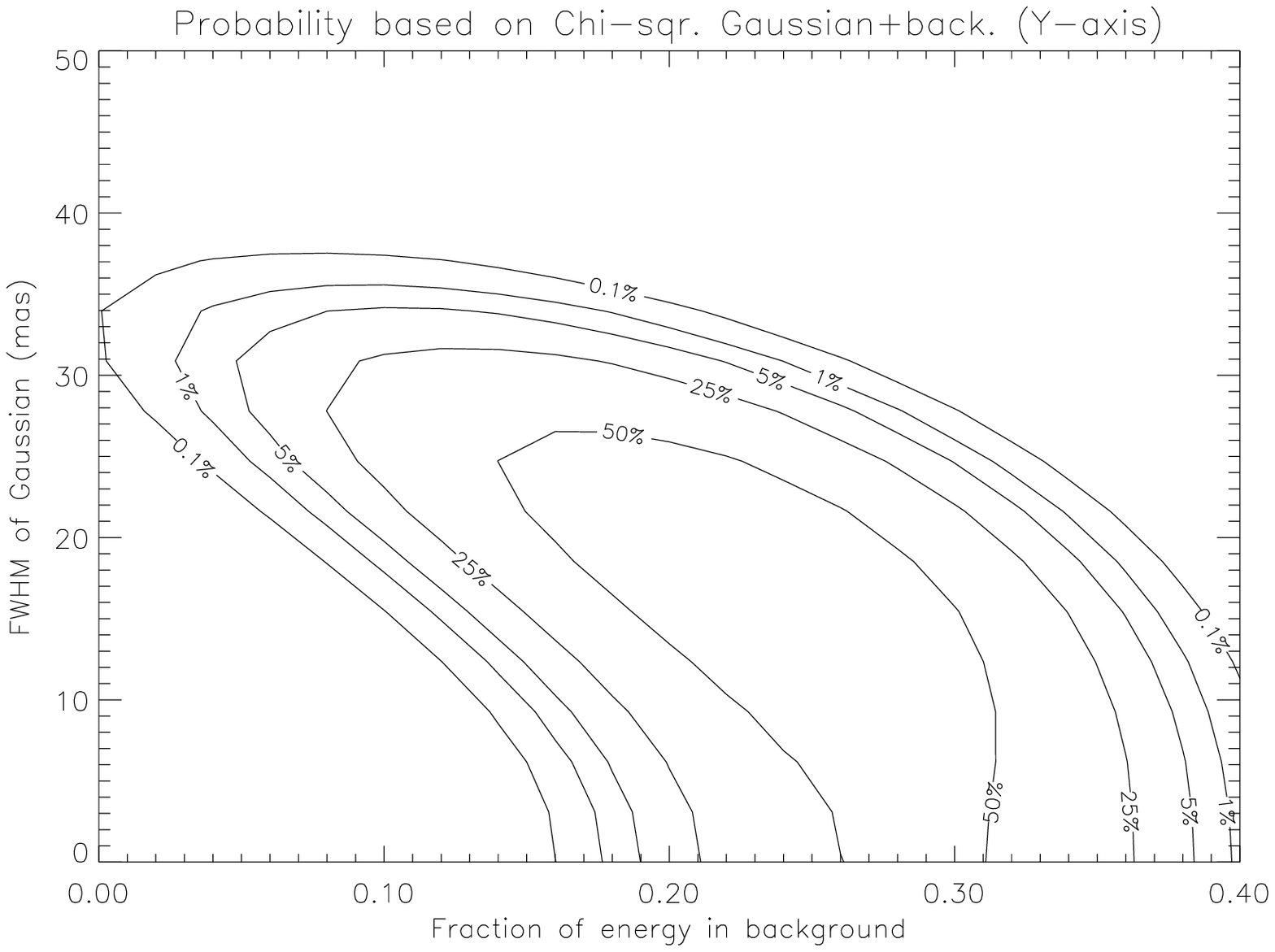}
\figcaption{Confidence limits of fits of a two parameter
set of models of a Gaussian and a flat background to the X and Y direction
S-curves for 3C279 dataset f0wj0602m.}
\label{Figure 2}
\end{figure}

\begin{figure}
\plotone{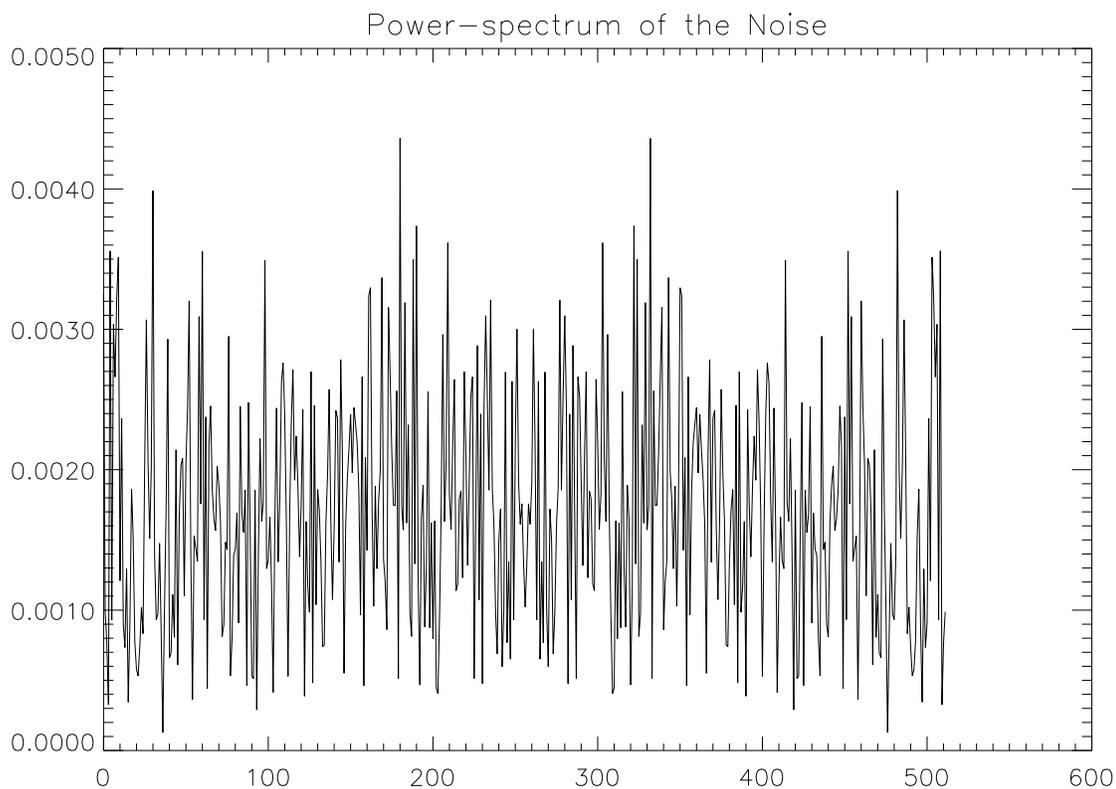}
\figcaption{Power-spectrum of the noise in a typical S-curve}
\label{Figure 3}
\end{figure}

\begin{figure}[t]
\plottwo{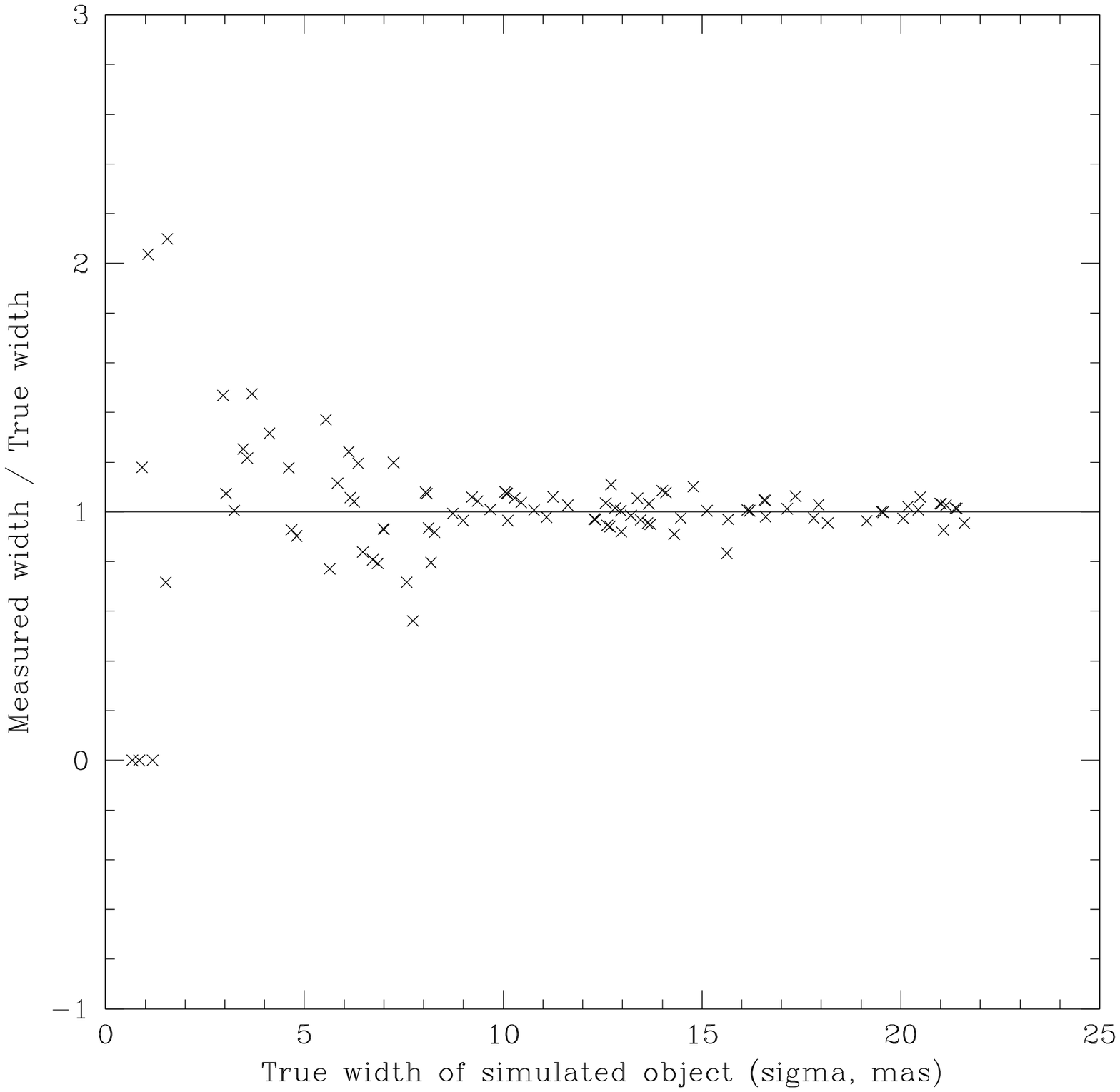}{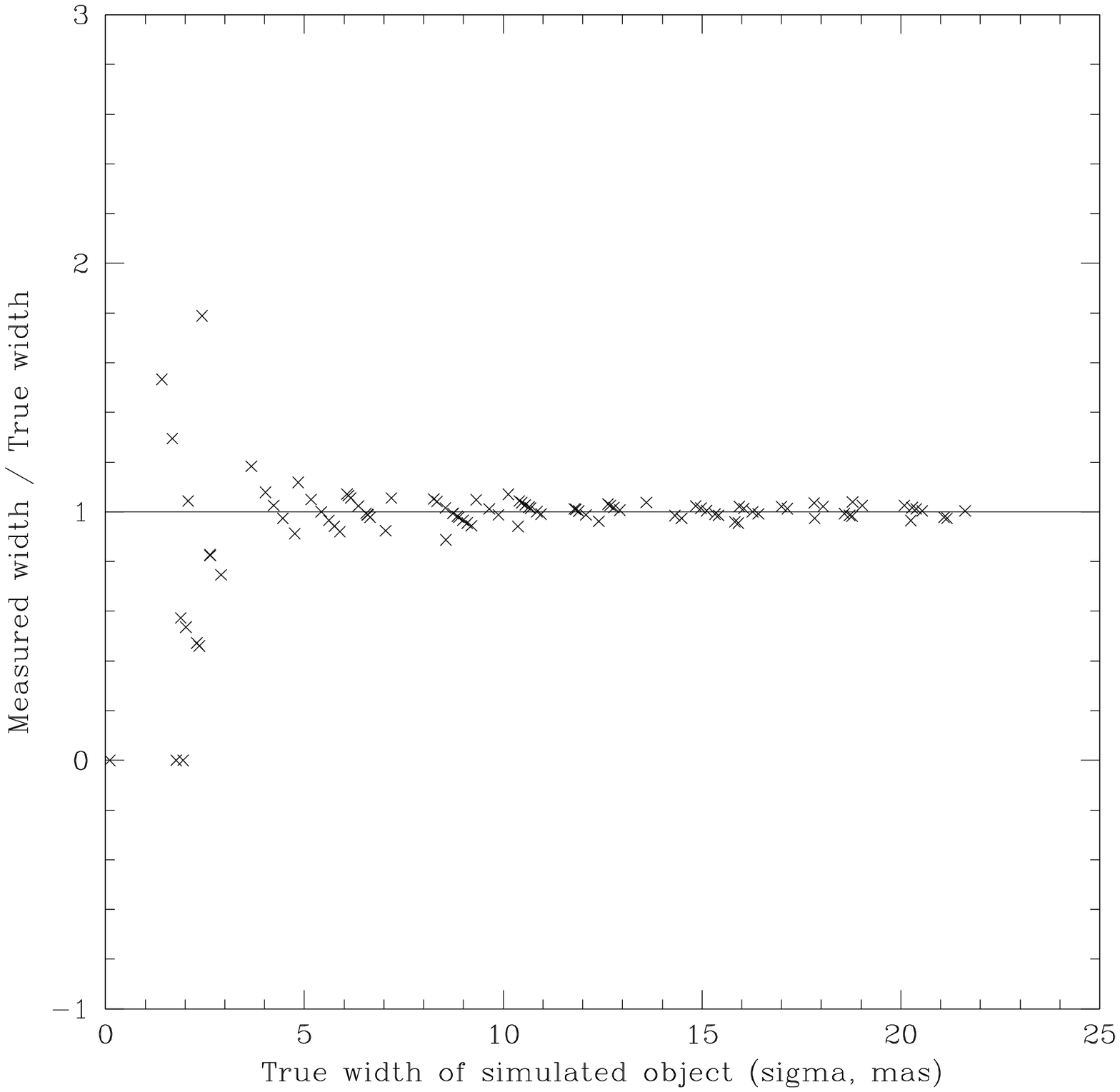}
\figcaption{Ratio of measured to true object extension as a function of
object width for simulations at two different noise levels. The
data for the left-hand plot had a noise level of 0.05 and the
right-hand one 0.01. The simulations are described further
in the text.}
\label{Figure 4}
\end{figure}

\begin{figure}[t]
\plottwo{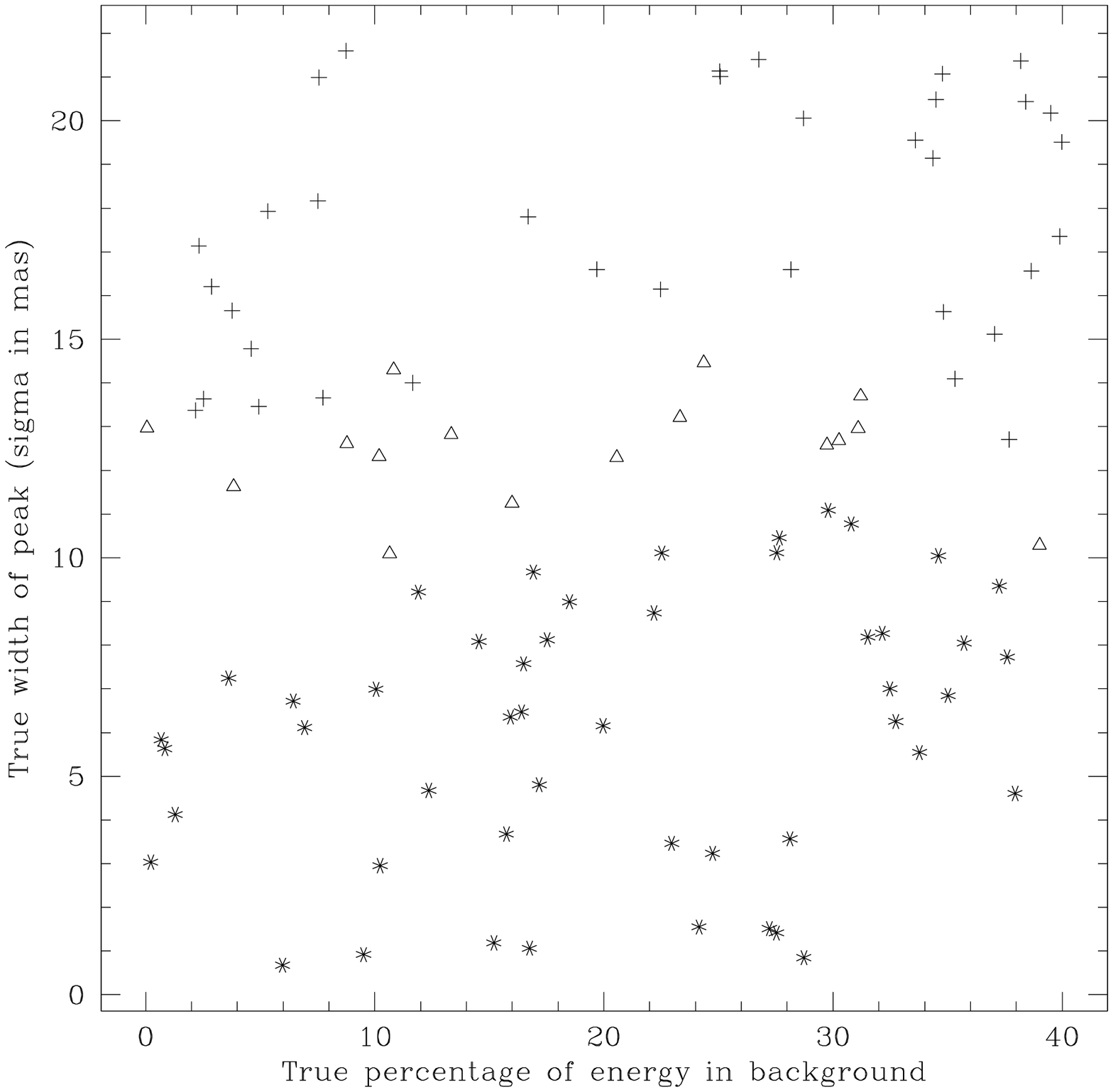}{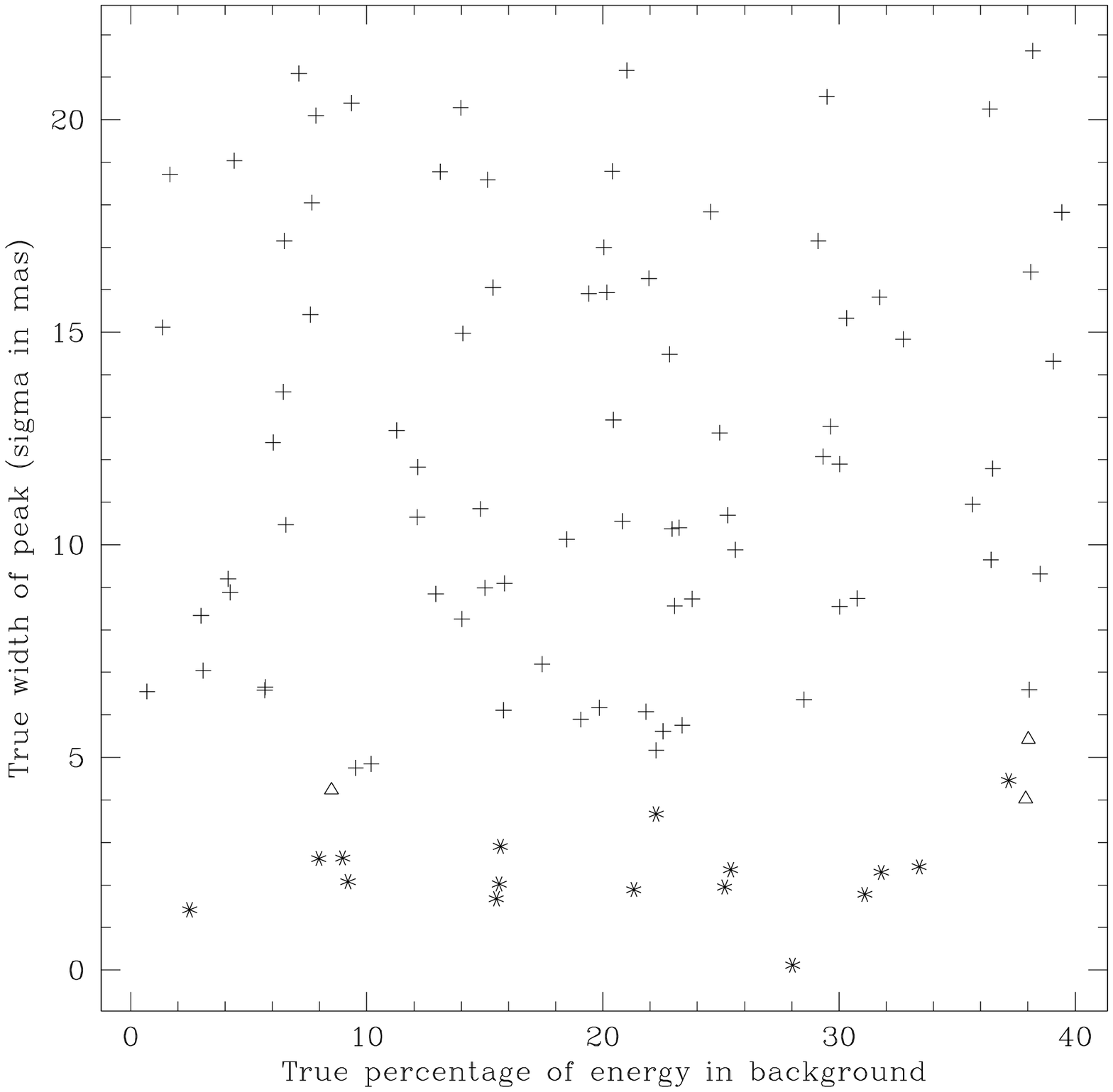}
\figcaption{Simulated S-curves measured and classified by whether they are
significantly non-point-like or not. The pluses represent fits which are
non-point-like at the 99\% confidence level. The plot at the left is
for a noise level of 0.05 and at the right 0.01. See the text for more
details.}
\label{Figure 5}
\end{figure}

\begin{figure}[t]
\plottwo{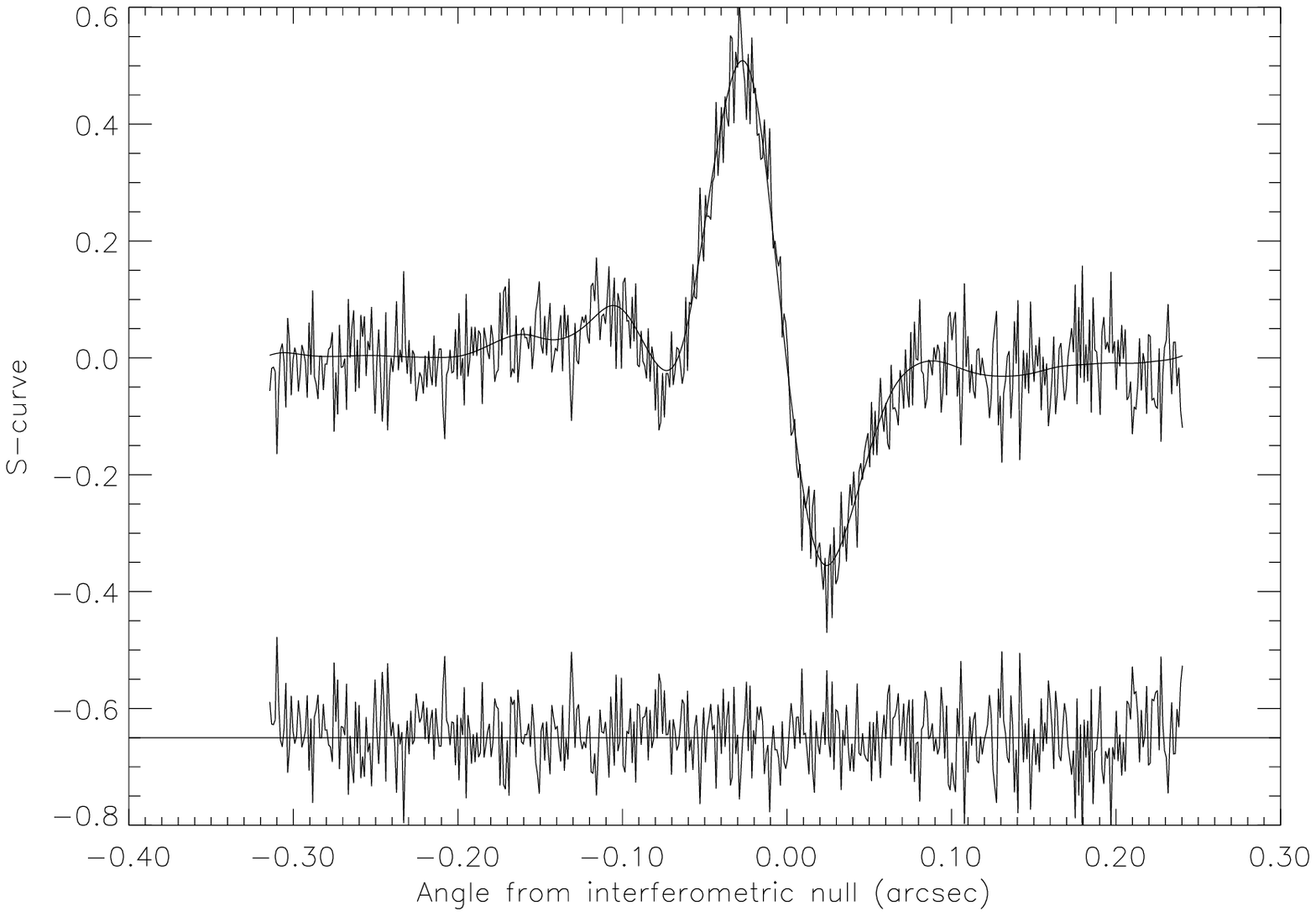}{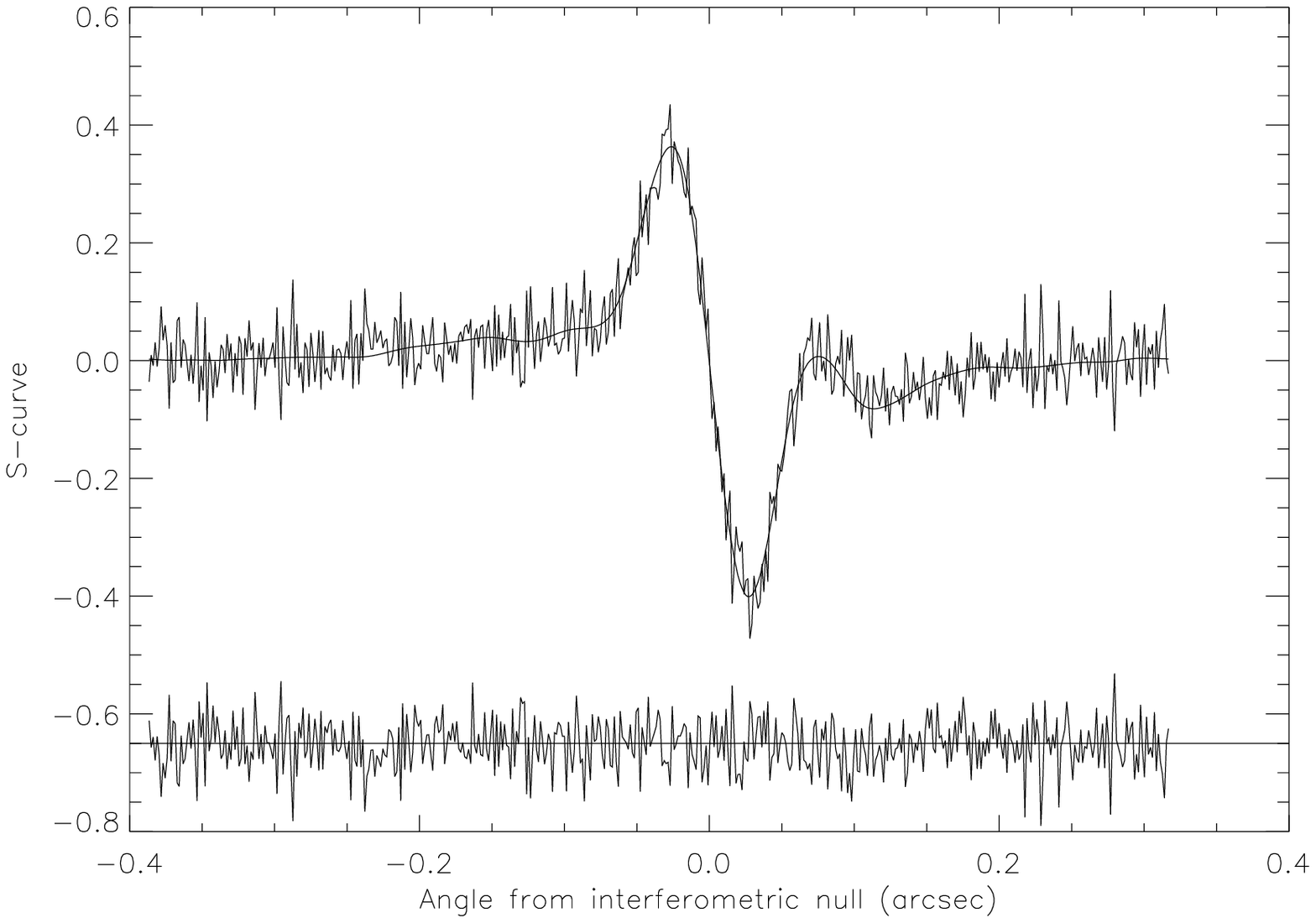}
\figcaption{Best fit to 3C279, 1992 April 2, X and Y S-curves}
\label{Figure 6}
\end{figure}

\begin{figure}[t]
\plottwo{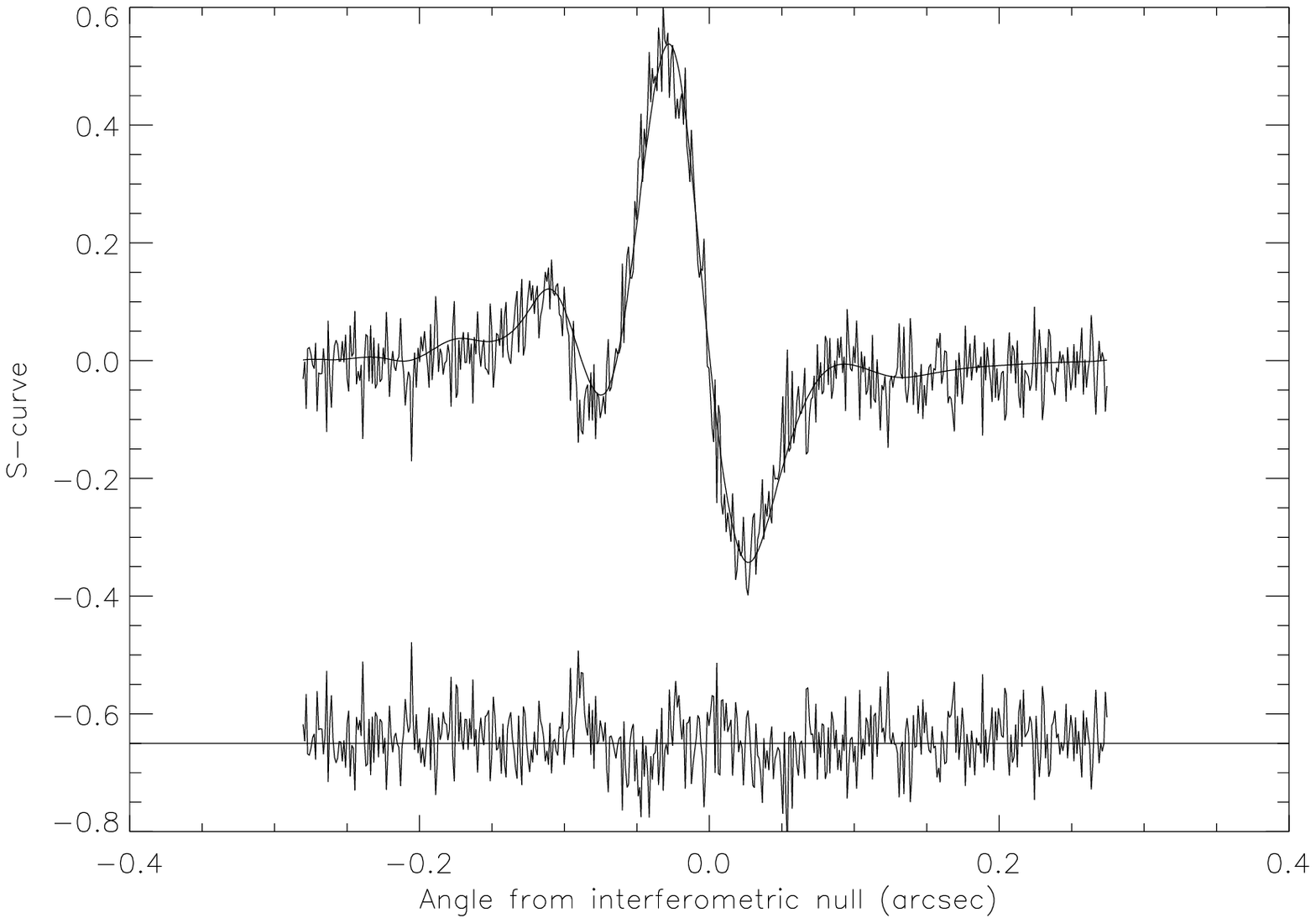}{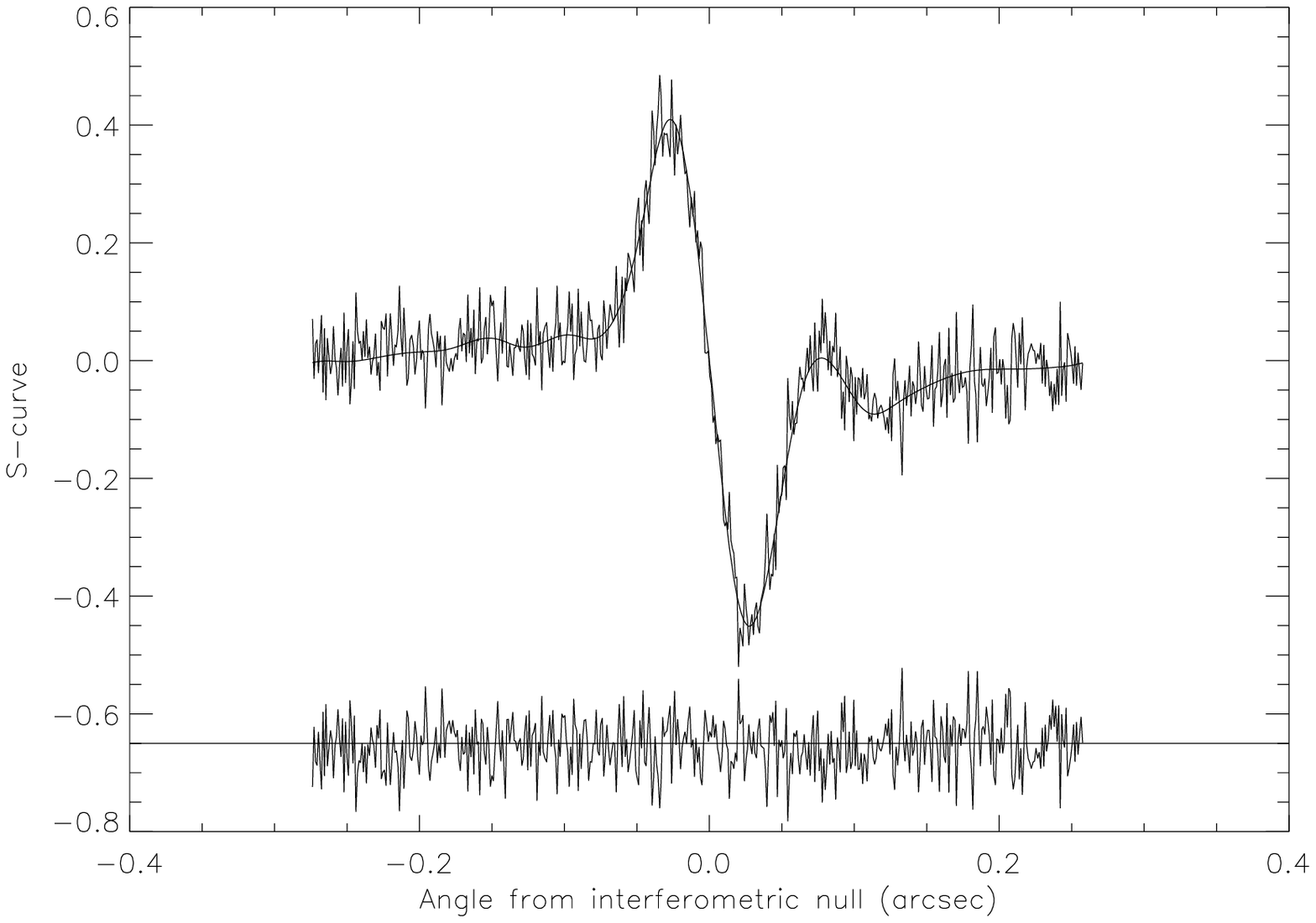}
\figcaption{Best fit to 3C279, 1995 April 10, X and Y S-curves}
\label{Figure 7}
\end{figure}

\begin{figure}[t]
\plottwo{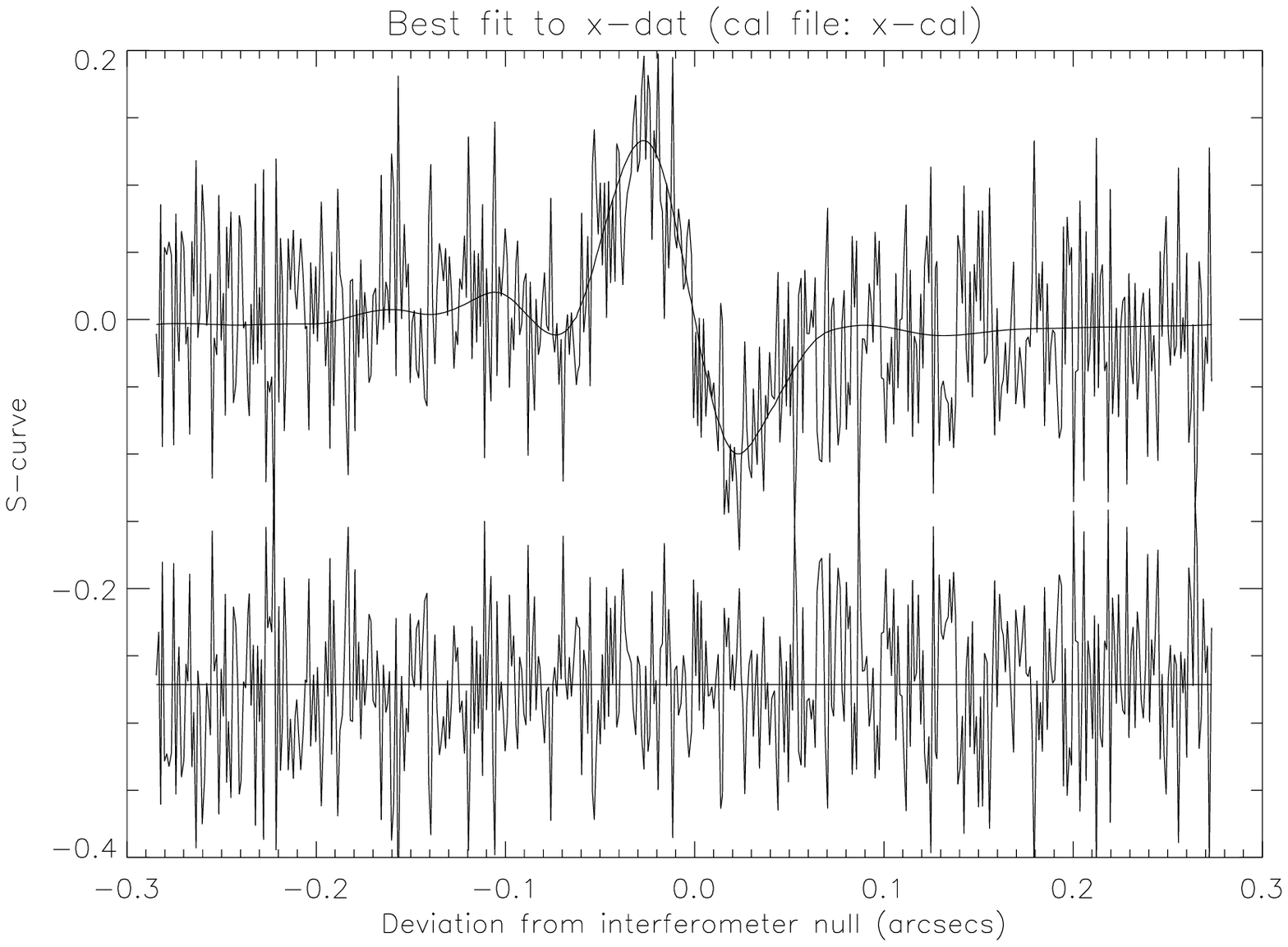}{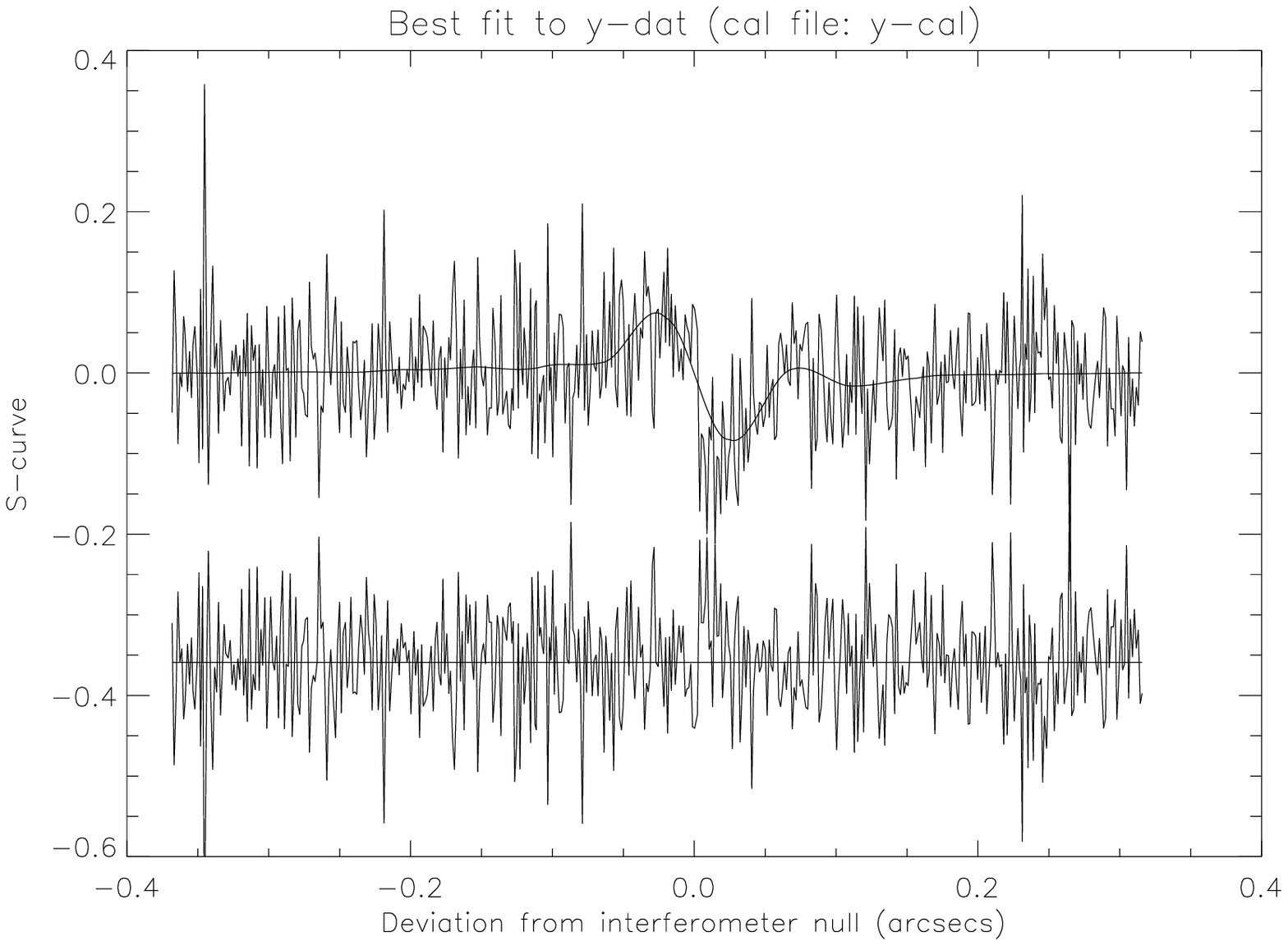}
\figcaption{Best fits to NGC1275 X and Y S-curves}
\label{Figure 8}
\end{figure}

\begin{figure}
\plottwo{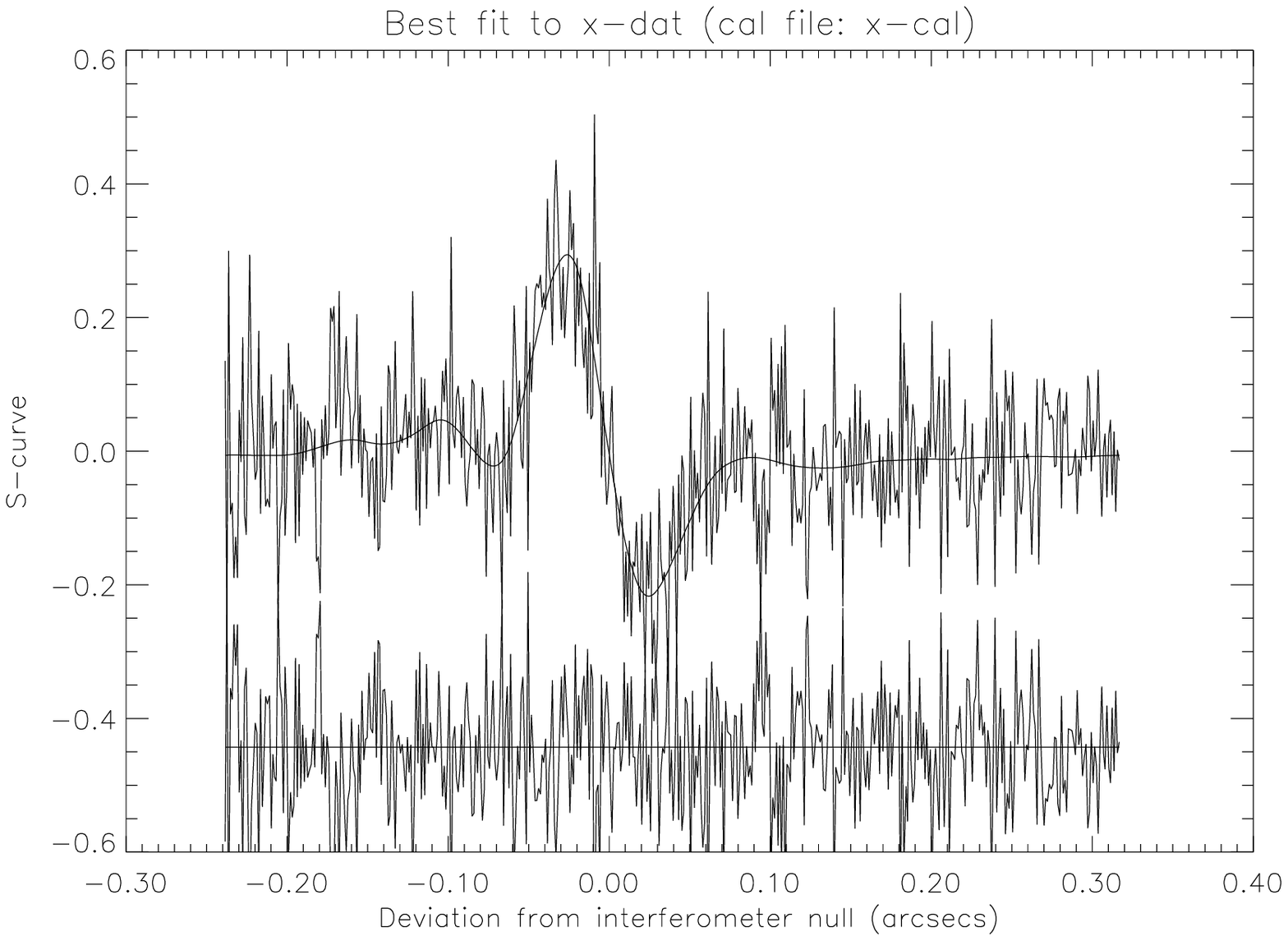}{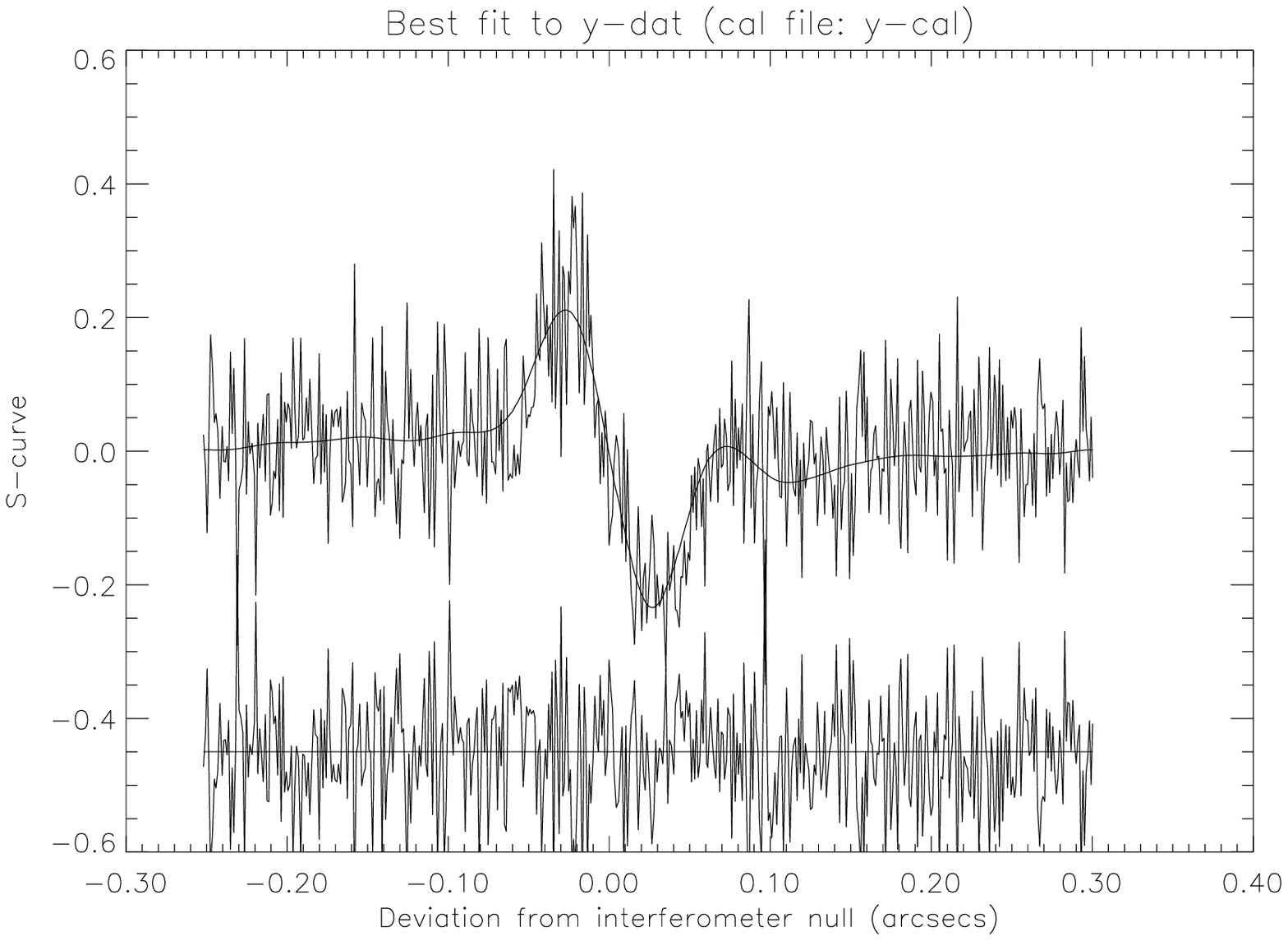}
\figcaption{Best fit to 3C345, 1993 April 12, X and Y S-curves}
\label{Figure 9}
\end{figure}

\begin{figure}
\plottwo{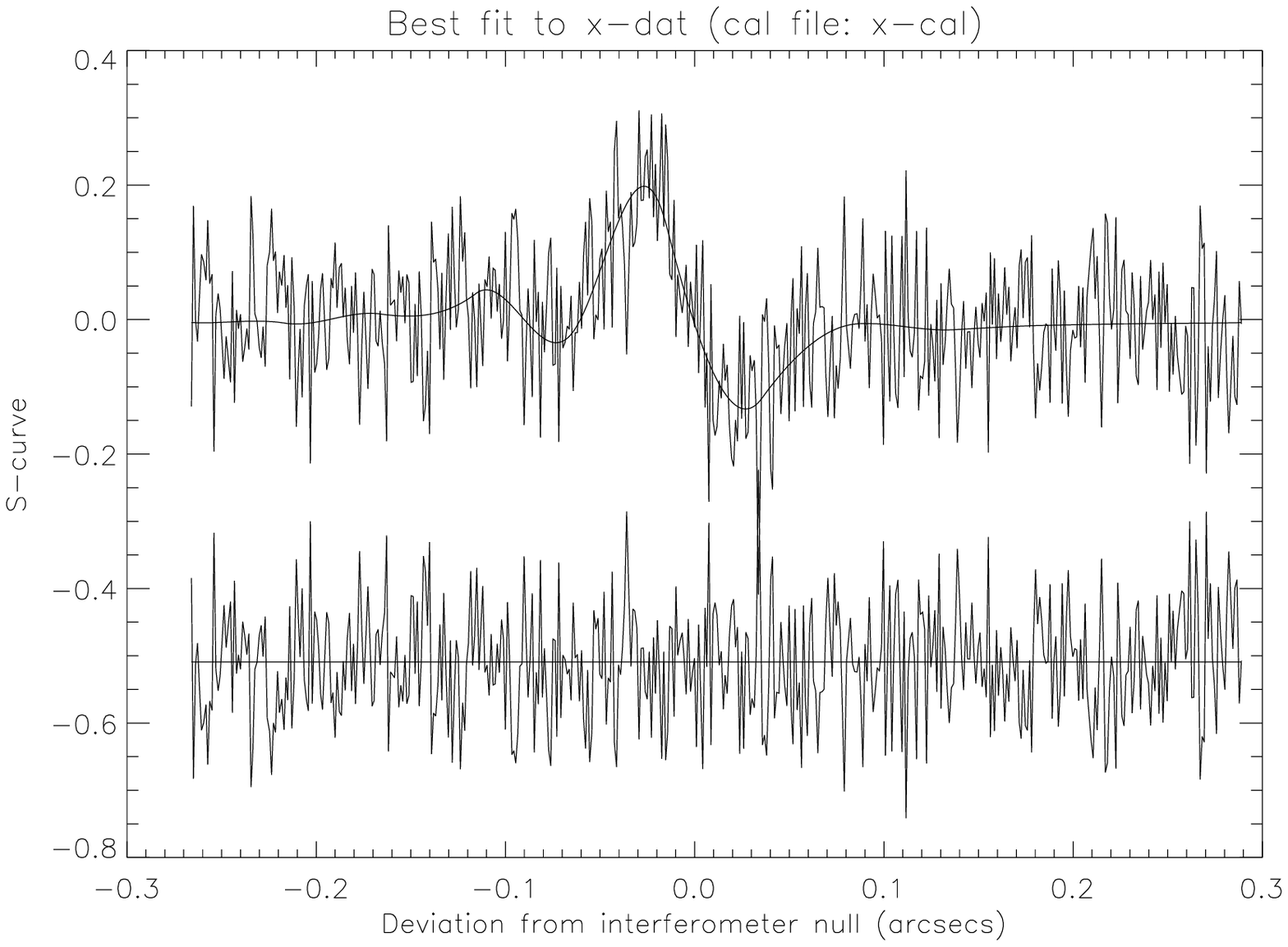}{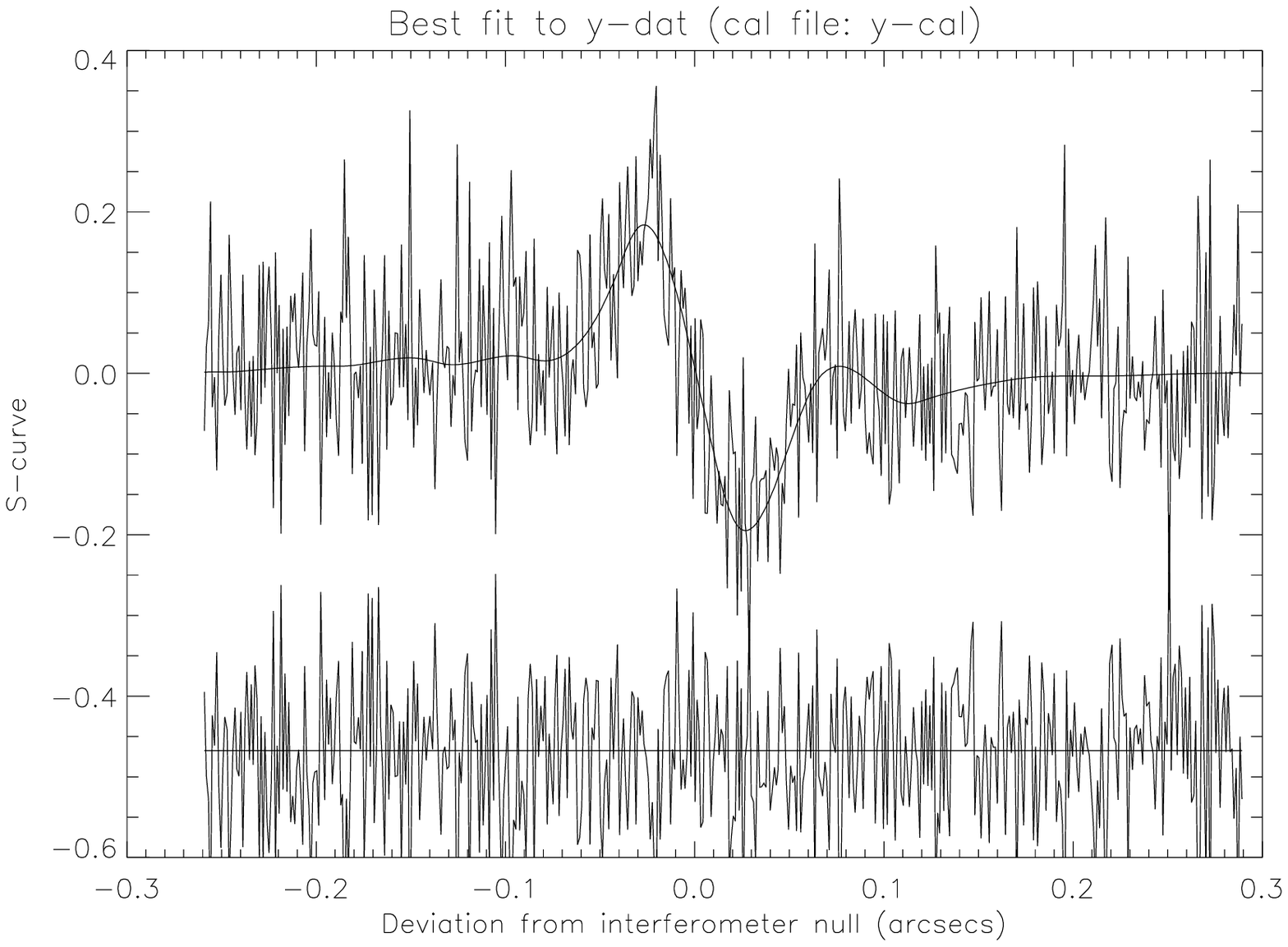}
\figcaption{Best fit to 3C345, 1995 February 22, X and Y S-curves}
\label{Figure 10}
\end{figure}

\begin{figure}
\plottwo{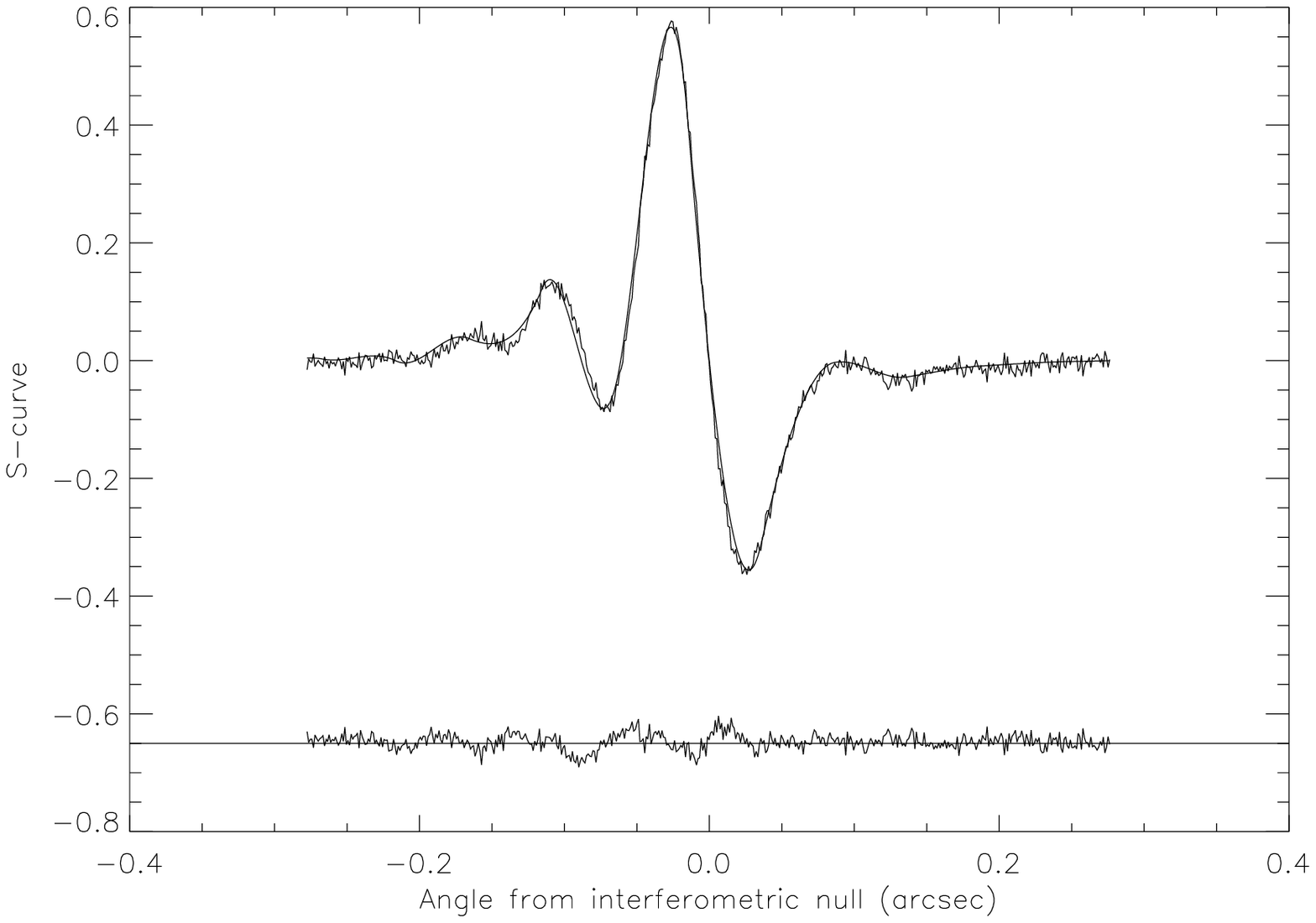}{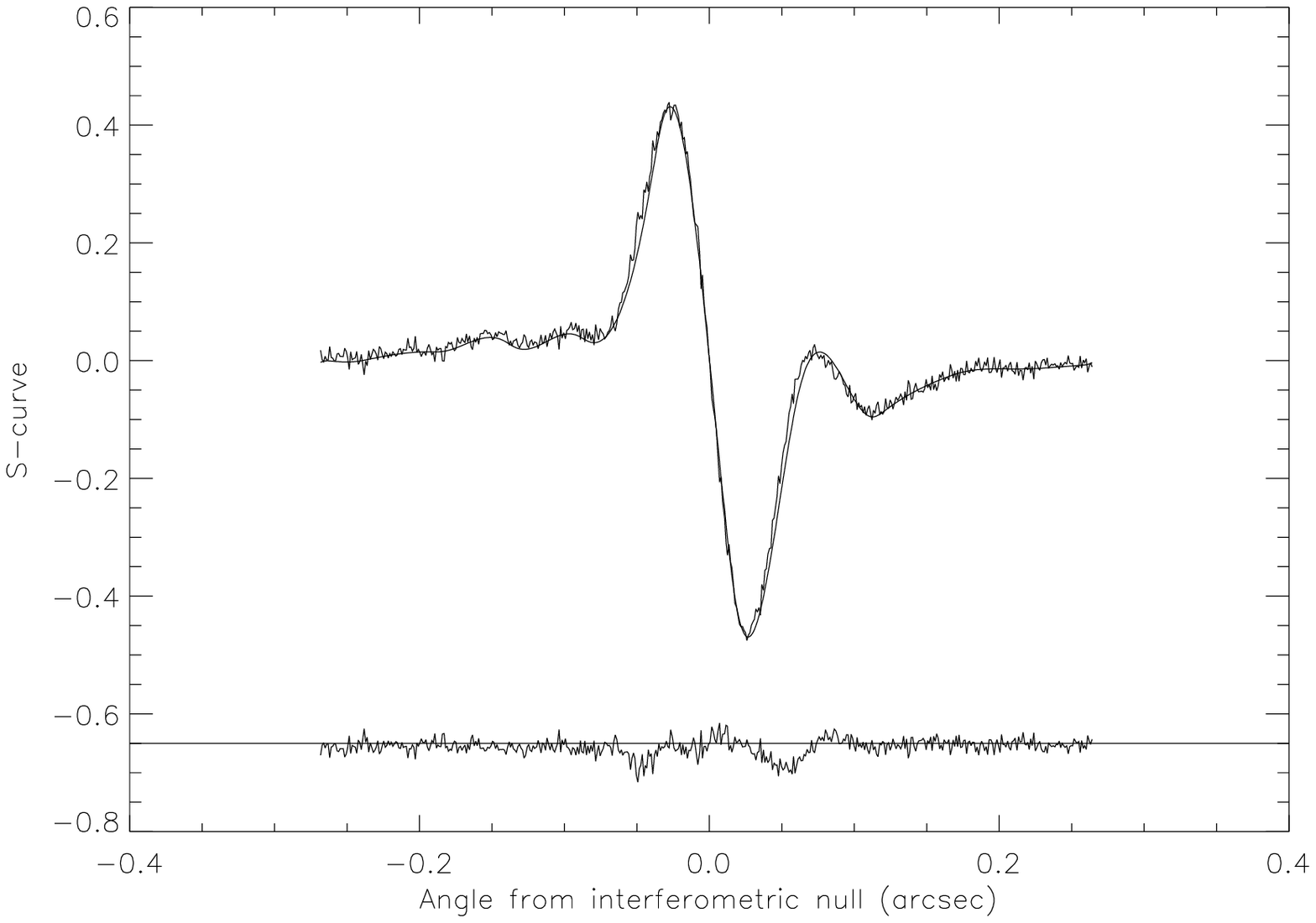}
\figcaption{Best fit to NGC4151 X and Y S-curves}
\label{Figure 11}
\end{figure}

\begin{figure}
\plottwo{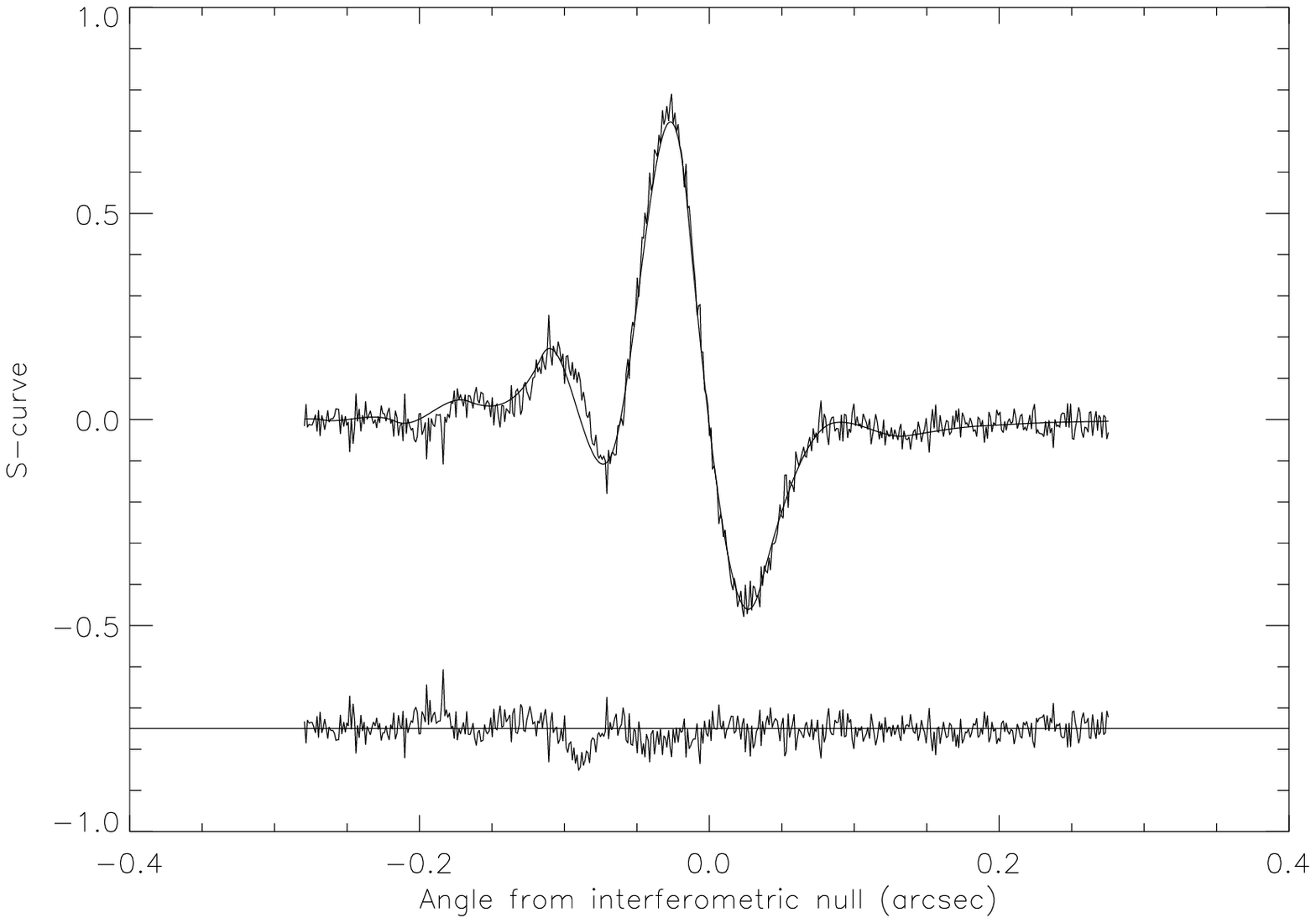}{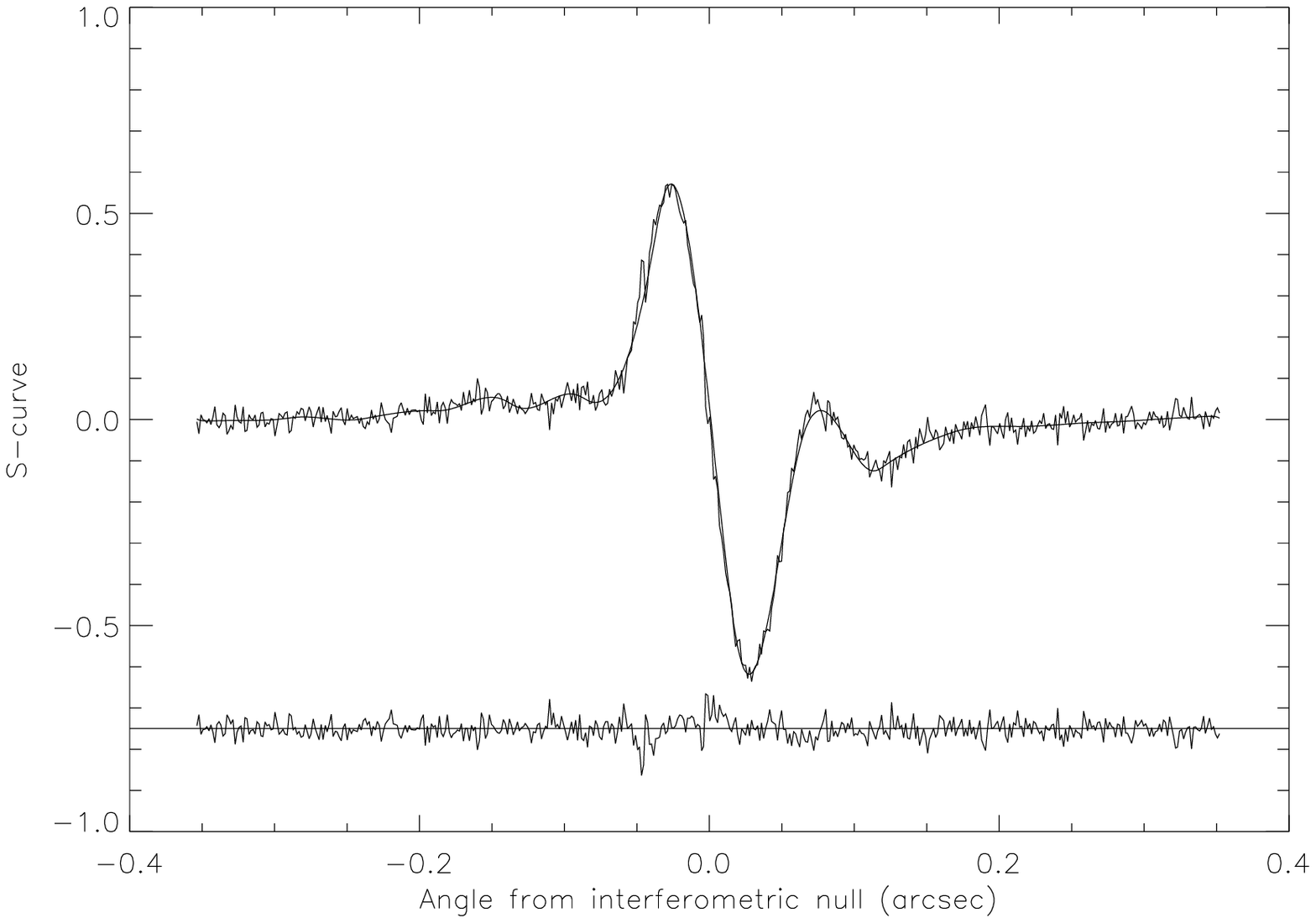}
\figcaption{Best fit to 3C273 X and Y S-curves}
\label{Figure 12}
\end{figure}

\begin{figure}
\plottwo{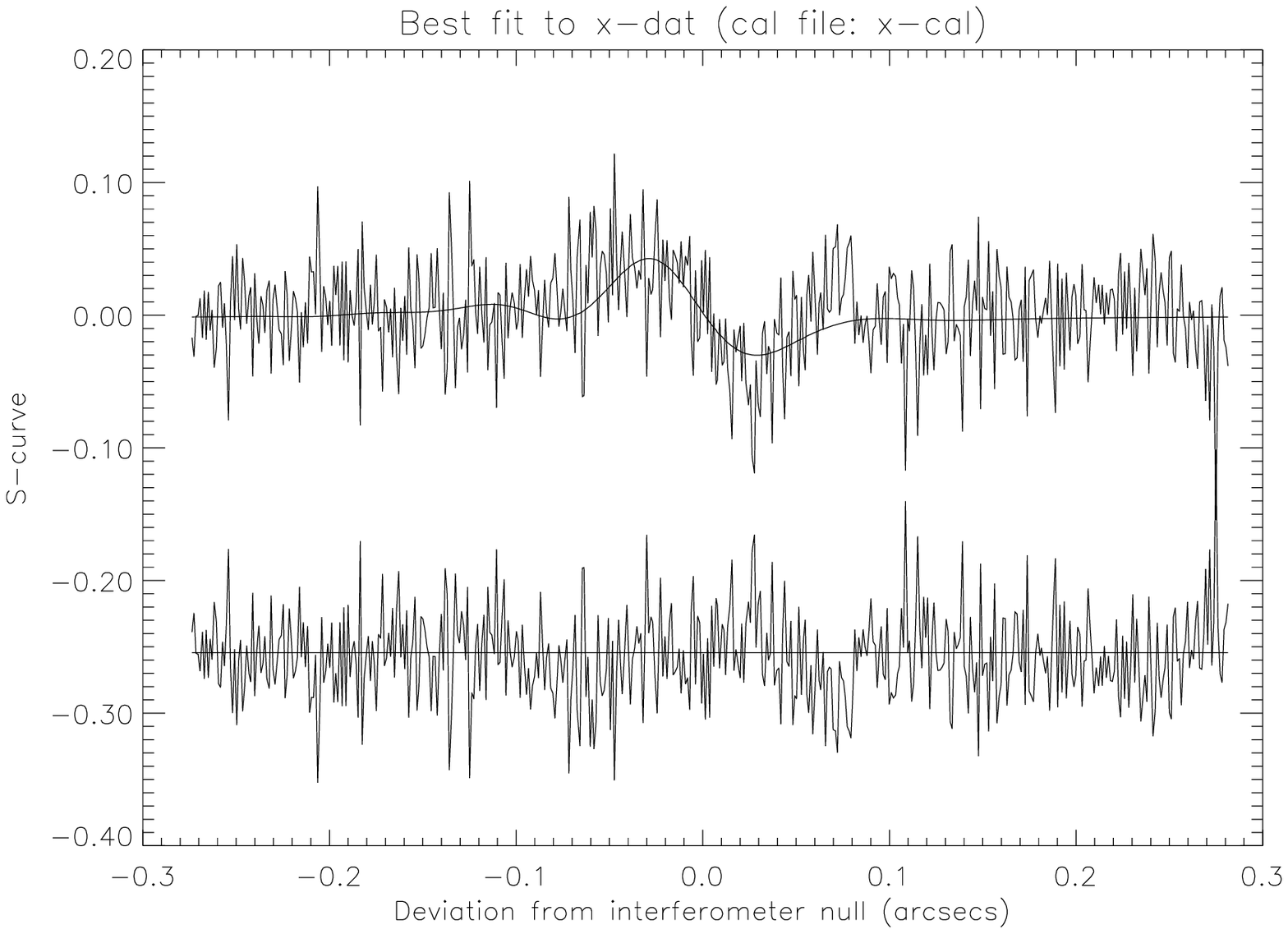}{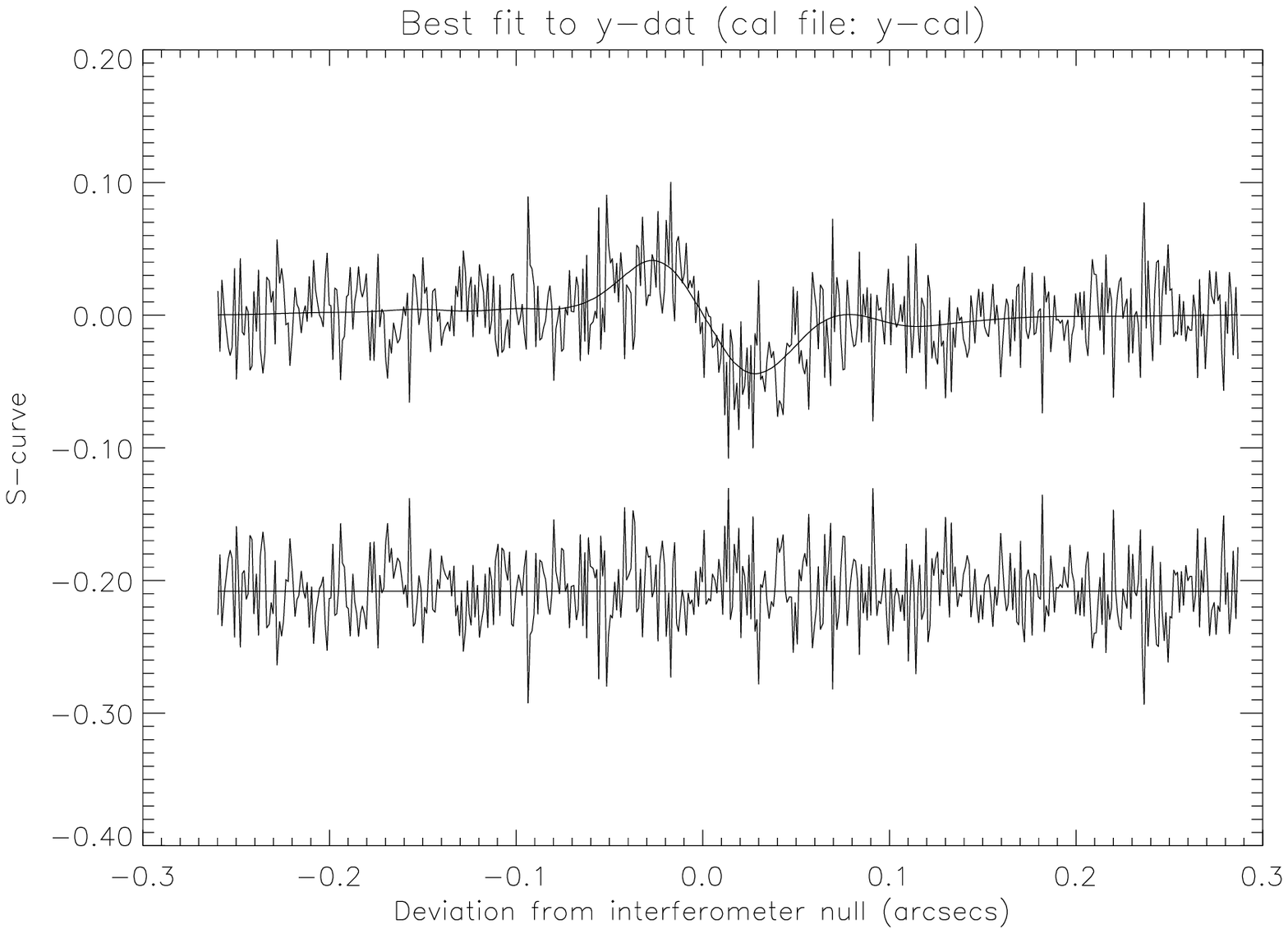}
\figcaption{Best fit to M87 X and Y S-curves}
\label{Figure 13}
\end{figure}

\clearpage

\begin{deluxetable}{cccccc}
\tablenum{1}
\tablewidth{0pt}
\tablecaption{Observed FGS AGN Datasets}
\tablehead{ \colhead{Object} & \colhead{Dataset} & \colhead{Observation}
 & \colhead{Exposure} & \colhead{Signal} &
\colhead{Calibration} \\
Name & Name & Date & Time (s) & Strength & Dataset}

\startdata
3C279 & f0wj0602m & 1992 Apr  2 & 1575 & strong  & f0v30302m \nl
NGC1275  & f0wj0203m & 1992 Oct 27 & 1508 & v.weak & f0v30702m \nl
3C345 & f0wj0702m & 1993 Apr 12 & 1590 & weak  & f0v30a02m \nl
NGC4151 & f2m40201m & 1995 Jan 29 & 2596 & v.strong & f2vc0201m \nl
3C345 & f2m40501m & 1995 Feb 22 & 1811 & weak  & f2vc0201m \nl
3C273 & f2m40302m & 1995 Mar 21 & 1096 & v.strong & f2vc0201m \nl
3C279 & f2m40401m & 1995 Apr 10 & 1811 & strong  & f2vc0201m \nl
M87   & f2m40101m & 1995 May 12 & 2596 & v.weak & f2vc0201m \nl
\enddata
\end{deluxetable}

\begin{deluxetable}{cccc}
\tablenum{2}
\tablewidth{0pt}
\tablecaption{List of calibration datasets of reference stellar sources}
\tablehead{ \colhead{Star} & \colhead{Dataset} & \colhead{Observation Date}
 & \colhead{Color $(B-V)$} }
\startdata
Upgren 69 & f0v30302m & 1992 Apr 6 & +0.5 \nl
Upgren 69 & f0v30702m & 1992 Sep 1 & +0.5 \nl
Upgren 69 & f0v30a02m & 1993 Apr 5 & +0.5 \nl
Upgren 69 & f2vc0201m & 1995 Sep 12 & +0.5 \nl
Upgren 69 & f2cs0502m & 1994 Jun 12 & +0.5 \nl
Upgren 69 & f3h80201m & 1996 Oct 23 & +0.5 \nl
Upgren 69 & f3h80501m & 1997 Jan 24 & +0.5 \nl
Upgren 69 & f3h85501m & 1997 Mar 30 & +0.5 \nl
LAT-COL-1B & f2vu0401m & 1996 Jan 9 & +0.18 \nl
\enddata
\end{deluxetable}

\begin{deluxetable}{ccccccc}
\tablenum{3}
\tablewidth{0pt}
\tablecaption{Fits to 3C273 S-curves using different calibration curves}
\tablehead{ \colhead{Calibration Dataset} &   \multicolumn{3}{c}{X}
& \multicolumn{3}{c}{Y}  \\
 & Background & $\sigma$ (mas) & $\chi^2_{red}$ & Background  &
$\sigma$ (mas) & $\chi^2_{red}$ }
\startdata
f2cs0502m &  0\% & 0 & 2.50 &     0\% &  0 & 2.29\nl
f2vc0201m &  0\% & 1 & 1.67 &     0\% &  3 & 1.47\nl
f3h80201m &  0\% & 0 & 1.26 &     0\% &  3 & 1.62\nl
f3h80501m &  0\% & 0 & 2.19 &     2\% &  1 & 1.45\nl
f3h85501m &  2\% & 0 & 1.86 &     0\% &  4 & 1.66\nl
f2vu0401m &  0\% & 0 & 1.43 &     0\% &  4 & 1.78\nl
\enddata
\end{deluxetable}

\begin{deluxetable}{cccccccc}
\tablenum{4}
\tablewidth{0pt}
\tablecaption{Results of two-parameter fits to S-curves}
\tablehead{ \colhead{Dataset} & \colhead{Object} & \multicolumn{3}{c}{X}
& \multicolumn{3}{c}{Y}  \\
Name & Name & Background & $\sigma$ (mas) & $\chi^2_{red}$ & Background  &
$\sigma$ (mas) & $\chi^2_{red}$}
\startdata
f0wj0602m& 3C279& 24\% & 5.4 & 1.10 &    28\% &  7.2 & 1.03\nl
f0wj0203m& NGC1275 & 80\% & 5.5 & 0.83 &    86\% &  0.0 & 0.80\nl
f0wj0702m& 3C345& 56\% & 5.4 & 0.91 &    58\% &  7.6 & 1.18\nl
f2m40201m& NGC4151& 22\% & 0.0 & 1.74 &    24\% &  3.1 & 2.6\nl
f2m40501m& 3C345& 72\% & 0.0 & 0.99 &    68\% &  3.2 & 0.91\nl
f2m40302m& 3C273& 0\% & 0.0 & 1.43 &     0\% &  2.1 & 1.50\nl
f2m40401m& 3C279& 18\% & 7.6 & 1.23 &    22\% &  7.3 & 0.91\nl
f2m40101m& M87  & 92\% & 13. & 1.20 &    92\% &  8.5 & 1.03\nl
\enddata
\end{deluxetable}

\begin{deluxetable}{cccccc}
\tablenum{5}
\tablewidth{0pt}
\tablecaption{Limits for physical extent and luminosity density of emitting
regions measured with FGS3. The performance of the pre-refurbishment FGS is
limited by systematics on the brighter objects.}
\tablehead{ \colhead{Object} & \colhead{z} & \colhead{Distance}
 & \colhead{Assumed} & \colhead{Upper Limit Radius} & \colhead{Lower Limit of
Luminosity} \\
Name &       & Modulus    & V    & of emitting region  &  Density
($L_\odot pc^{-3}$)}
\startdata
3C279 & 0.536     & 42.0   & 15  & 12mas = 73pc &  $6.9 \times 10^6$ \nl
NGC4151 & 0.0033 & 30.9   & 13  & 9mas = 0.68pc &  $ 9.1 \times 10^8$ \nl
3C273 & 0.158     & 39.3   & 12.8 & 9mas = 26pc & $ 4.7 \times 10^7$ \nl
\enddata
\end{deluxetable}

\begin{deluxetable}{cccc}
\tablenum{6}
\tablewidth{0pt}
\tablecaption{Estimated limits for physical extent and luminosity density
of emitting regions which should be attainable with the post-refurbishment
FGS1R assuming that the performance will be limited primarily by photon
statistics rather than systematics of the instrument.}
\tablehead{ \colhead{Object} &  \colhead{Upper Limit Radius} &
\colhead{Lower Limit of Luminosity} \\
Name &    of emitting region  &  Density ($L_\odot pc^{-3}$)}
\startdata
3C279 & 12mas = 73pc &  $6.9 \times 10^6$ \\
NGC4151 & 5mas = 0.38pc &  $ 5.3 \times 10^9$ \\
3C273 & 7mas = 20pc & $ 1.0 \times 10^8$ \\
\enddata
\end{deluxetable}


\begin{references}

\reference{baa92} B\aa \aa th, L.B., Rogers, A.E.E., Inoue, M., 
Padin,S., Wright, M.C.H., Zensus, A., Kus, A.J., Backer, D.C., 
Booth, R.S., \& Carlstrom, J.E. 1992, \aap,  257, 31

\reference{cas97} Casidy, I, \& Raine, D.\ J. 1997, \mnras, 322, 400

\reference{dep83} de Pater, I. \& Perley, R.A. 1983, \apj, 273, 64

\reference{fra92} Franz, O.G., Wasserman, L.H., Nelan, E., Lattanzi, 
M.G., Bucciarelli, B. \& Taff, L.G. 1992, \aj, 103, 190

\reference{her92} Hershey, J.L. 1992, \pasp, 104, 592

\reference{lat94} Lattanzi, M.G., Hershey, J.L., Burg, R., Taff, L.G., 
Holfeltz, S.T., Bucciarelli, B., Evans, I.N., Gilmozzi, R., Pringle, J. 
\& Walborn, N.R. 1994, \apj, 427, L21

\reference{lat97} Lattanzi, M.G., Munari, U., Whitelock, P.A. \& 
Feast, M.W. 1997, \apj, 485, 328

\reference{san65} Sandage, A. \& Wyndham, J.D. 1965, \apj, 141, 328
\end{references}
\end{document}